\documentclass[twocolumn]{aastex631}
\usepackage{showyourwork}
\usepackage{amsmath}

\newcommand{\kms}{${\rm km~s^{-1}}$}

\newcommand{\CMC}{\texttt{CMC}}
\newcommand{\CMCcat}{\texttt{CMC Cluster Catalog}}
\newcommand{\fewbody}{\texttt{Fewbody}}

\shorttitle{Runaway \& hypervelocity stars from globular clusters}
\shortauthors{Cabrera \& Rodriguez}

\begin{document}

% Title
\title{Runaway and Hypervelocity Stars from Compact Object Encounters in Globular Clusters}

% Author list
\author[0000-0002-1270-7666]{Tom\'as Cabrera}
\affiliation{McWilliams Center for Cosmology,
    Department of Physics,
    Carnegie Mellon University,
    5000 Forbes Avenue, Pittsburgh, PA 15213
}

\author[0000-0003-4175-8881]{Carl L. Rodriguez}
\affiliation{McWilliams Center for Cosmology,
    Department of Physics,
    Carnegie Mellon University,
    5000 Forbes Avenue, Pittsburgh, PA 15213
}
\affiliation{
    Department of Physics and Astronomy,
    University of North Carolina at Chapel Hill,
    120 E. Cameron Ave, Chapel Hill, NC, 27599, USA
}

% Abstract with filler text
\begin{abstract}
    The dense environments in the cores of globular clusters (GCs) facilitate many strong dynamical encounters among stellar objects.
    These encounters have been shown capable of ejecting stars from the host GC, whereupon they become runaway stars, or hypervelocity stars if unbound to the galactic potential.
 	We study high speed stellar ejecta originating from GCs by using Monte Carlo $N$-body models, in particular focusing on binary-single encounters involving compact objects.
 	We pair our model-discriminated populations with observational catalogs of Milky Way GCs to compose a present-day galactic population of stellar ejecta.
 	We find that these kinds of encounters can accelerate stars to velocities in excess of 2000 \kms, to speeds beyond the previously predicted limits for ejecta from star-only encounters and in the same regime of galactic center ejections.
	However, the same ejections can only account for 1.5-20\% of the total population of stellar runaways, and only 0.0001-1\% of hypervelocity stars, with similar relative rates found for runaway white dwarfs.
	We also provide credible regions for ejecta from 149 Milky Way GCs, which we hope will be useful as supplementary evidence when pairing runaway stars with origin GCs.
\end{abstract}

% Main body with filler text
\section{Introduction}
\label{sec:intro}

Hypervelocity stars (HVSs) are stars that have been accelerated beyond the local galactic escape speed, that is, to the point of becoming unbound from the galactic potential.
Theorized by \citealt{1988Natur.331..687H}, the classical origin of HVSs is the dynamical disruption of a stellar binary by a super massive black hole (SMBH); this ``Hills mechanism" is believed to be capable of accelerating stars to speeds up to 4000 km s$^{-1}$, and naturally makes HVSs a potential probe of the galactic center.
Since the first HVS discovery \citep{2005ApJ...622L..33B}, a handful of candidate objects have been identified in the Milky Way (MW) (e.g. \citealt{2014ApJ...787...89B}), including the object S5-HVS1, with a measured speed of $\sim$1700 km s$^{-1}$ \citep{2020MNRAS.491.2465K}.
While S5-HVS1, along with several others, is well-explained by accepting an origin from the galactic center, for which there is overwhelming evidence for an SMBH (\citealt{1998ApJ...509..678G}, \citealt{2018A&A...615L..15G}, \citealt{2022ApJ...930L..12E}), recent studies have found examples of high-velocity stars that are not so easily read \citep[e.g.][]{2018MNRAS.479.2789B, 2019MNRAS.483.2007E, 2021A&A...646L...4I}.
Two consistent results from these lines of work are that many stars previously classified as HVSs are nonetheless bound to the galactic potential, and many of these ``runaways" are more likely to have been ejected from the disk or a satellite as opposed to the galactic center.
Altogether, the identification of systems of origin for HVSs and runaway stars is a necessary threshold problem to unlocking the full potential of observations of these objects.

Alternate acceleration mechanisms capable of producing high-velocity stars include the binary supernova and dynamical ejection scenarios (BSS \citep{1961BAN....15..265B} and DES \citep{1967BOTT....4...86P}, respectively).
In the BSS, the more massive primary of a stellar binary undergoes a supernova, which subsequently accelerates the companion; this scenario has been predicted to accelerate stars to speeds of a few hundred \kms \citep{2019A&A...624A..66R, 2022MNRAS.tmp.3236I}, with exceptionally light companions potentially receiving velocities in excess of 1000 \kms \citep{2015MNRAS.448L...6T}.
The DES concerns strong gravitational interactions among 3-4 stellar objects, and has been associated with higher speed limits: \citet{1991AJ....101..562L} studied these encounters with numerical methods, and found that the upper limit for products of these encounters was approximately the escape speed from the surface of the most massive star involved.
For a sun-like star, this escape speed is $\sim$620 \kms, while for a 60 $M_\odot$ late main-sequence star it is $\sim$1400 \kms.
\citet{2020ApJ...903...43D} found that for OB runaways in the Small Magellanic Cloud, the DES is expected to occur 2-3 times more frequently that the BSS.

Globular clusters (GCs) are obvious candidate matrices for both of these events due to their high stellar densities, but the DES might be exceptionally amplified due to the presence of BH subsystems in the centers of GCs.
Multi-band observations \citep{Maccarone2007a, Barnard2011, Chomiuk2013, Miller-Jones2015a} employing a variety of techniques \citep{Strader2012, Bahramian2016, Giesers2018} have been successful in finding tens of BH candidates in GCs, which are in themselves consistent with the presence of hundreds of BHs per cluster \citep{2018ApJ...855L..15K, Giesers2019}.
BHs play important roles in the macroscopic evolution of the GC \citep[e.g.][]{2020IAUS..351..357K} by dominating gravitational interactions in GC cores; this naturally places BHs in a privileged position when considering the DES in GCs, particularly when considering the maximum velocities attainable by this mechanism.

Several recent and ongoing studies have investigated these phenomena and the broader question of extra-tidal stars from GCs, in part encouraged by the continual improvement of astrometric measurements through efforts such as \textit{Gaia} \citep{2022arXiv220800211G}.
\citet{2023MNRAS.518.4249G} developed and used a particle spray code to model ejections from GC cores and demonstrate that kinematic cuts can be too aggressive when searching for ejecta, preferring the use of chemical abundances alone.
\citet{2023arXiv230105166F} presented a comprehensive catalog of extra-tidal features of MWGCs, produced by simulating tidal stripping of known MWGCs in the context of the MW potential.
Of particular proximity to our work, \citet{2022arXiv221116523W} used the same GC model catalog we employ to holistically examine stellar ejections from GCs and identify key mechanisms.
The latter two of of these works are explicitly presented as the first in a series of papers seeking to compose a more complete picture of their respective objects.

This work investigates the capacity of GCs to produce HVSs through the DES.
We focus on encounters between binary and single objects, as they are the most abundant kind of strong encounter, and specifically those that involve compact objects (COs, meaning BHs, a white dwarfs (WDs), and neutron stars (NSs)).
In \S\ref{sec:methods}, we describe our method of generating binary-single compact object (BSCO) ejecta populations from realistic simulations of GCs.
In \S\ref{sec:gcdyns}, we examine these populations as functions of GC parameters, and the relationship they have to GC evolution.
In \S\ref{sec:est_MW-like}, we combine our synthetic populations with observational MWGC catalogs to produce a MW-like population of ejecta, along with predicting rates and phase space distributions for these objects.
We conclude that GCs are possible generators of HVSs in all velocity regimes thus observed, beyond the previously established limit for star-only encounters, albeit the rate at which these objects are produced from GCs is significantly lower than that for the galactic center.

\section{Methods: Sampling \& Integrating BSCO encounters} \label{sec:methods}

The \CMCcat\ was generated using \CMC, a H\'enon-style $N$-body code for collisional stellar dynamics.  Developed over two decades \citep{2000ApJ...540..969J,2013ApJS..204...15P,Rodriguez2022}, \CMC\ relies upon the technique originally developed by \citet{1971Ap&SS..13..284H,1971Ap&SS..14..151H}, where the cumulative effect of two-body encounters is modeled as an ``effective'' encounter between neighboring particles (in a radially sorted, spherically symmetric  potential).  Because these neighboring particles are individual stars (or binaries), the H\'enon method allows detailed stellar and strong dynamical encounters to be considered as well. To that end \CMC~includes prescriptions for three-body binary formation from single BHs \citep{2013ApJ...763L..15M}, binary-single and binary-binary gravitational encounters using the \fewbody\ package \citep{2004MNRAS.352....1F,2007ApJ...658.1047F}, and galactic tidal fields \citep{2013MNRAS.429.2881C}.  The version of \CMC\ used to create the \CMCcat~\citep[which used identical physics to the public version described in][]{Rodriguez2022} treats single and binary stellar evolution for stars with  the \texttt{COSMIC} code for population synthesis \citep{2020ApJ...898...71B}.  \texttt{COSMIC} is based upon the original Binary Stellar Evolution (BSE) code \citep{2000MNRAS.315..543H,2002MNRAS.329..897H}, but with updated prescriptions for compact-object formation and massive star evolution; see \citet{2020ApJ...898...71B} and \citet{Rodriguez2022} for details.

When one of the two neighboring particles in a cluster is a binary, \CMC\ calculates whether to perform a strong dynamical encounter by calculating the probably $P_{\rm BS}$ for an encounter to occur within a single timestep $\Delta T$ as
\begin{equation}
P_{\rm BS} = n \Sigma w \Delta T~,
\label{eqn:pbs}
\end{equation}
where $n$ is the local density of stars, $w$ is the relative velocity between the neighboring star and binary, and $\Sigma$ is the cross section for encounters to occur, given by
\begin{equation}
\Sigma = \pi r_p^2 \left(1+\frac{2 G M}{r_p w^2}\right),
\label{eqn:sigma}
\end{equation}
where $M$ is the total mass of the system and $r_p$ is the radius within which a strong encounter is assumed to occur (equal to twice the binary semi-major axis by default).  
During each timestep, \CMC\ determines whether to perform a strong three-body encounter between a neighboring star and binary by computing the probability from \ref{eqn:pbs} and comparing it to a random variable drawn from [0,1], $X$.  If $X < P_{\rm BS}$, an encounter is performed with an impact parameter, $b$, selected from a distribution proportional to $P(b) \sim b~db$ out to a maximum integrated area set by \ref{eqn:sigma}.

At this point the encounter is handed off to \fewbody\ \citep{2004MNRAS.352....1F}, a small-$N$-body gravitational scattering code.
\fewbody\ randomly selects all remaining parameters, such as the phase of the binary and the orientation of the angular momentum and Runge-Lenz vectors; of course, this means that any encounter produced in a single \CMC\ integration is only a single realization of all the possible encounters that could have occurred in the cluster at that time.
For binary-single encounters, a hyperbolic encounter between the binary center-of-mass and the single object is initiated with infinite separation, and is analytically advanced until the tidal perturbation of the binary reaches a set threshold.
An 8$^{\rm th}$-order Runge-Kutta Prince-Dormand integrator then evolves the encounter, and the participating objects are classified as singles or into binaries at regular timestep intervals.
If the reduced encounter among the present single(s) and possible binary has a positive energy and the single(s)/binary are moving away from each other, then integration is terminated (termination also occurs if the encounter becomes analytic e.g. a merger results in a Keplerian interaction).

We implement additional features into \fewbody\ pertinent to the study of compact object dynamics.
We add additional parameters to track stellar object type, distinguishing between stars and compact objects.
We use these object types to decide whether to use \fewbody's default sticky-sphere collision criterion (which triggers a collision when the separation of the objects' centers is less than the sum of their radii) in the case of stellar interactions, or a tidal disruption criterion in the case of encounters involving NSs and BHs, which multiplies the threshold separation by a factor of $(m_{\rm CO} / m_{*})^{1/3}$ \citep[as was done in][]{Kremer2019}.
To the utility of this collision metric, we conservatively assign BHs radii of 5 times their Schwarzschild radii to group the more extreme encounters with mergers, both of which require more careful treatment to accurately compute.

To establish realistic binary-single encounter populations for clusters of various parameters, we take the 148 models from the \CMCcat\ and extract the initial conditions of all strong binary-single encounters that involve at least one luminous object (star, WD, or NS; to limit the focus to encounters that can produce observable ejecta) and one CO.
Over the entire catalog, our encounter sample makes up about half of all strong encounters that occur in the models, with each model contributing a few ten thousand BSCO encounters on average.
We realize each encounter in isolation with our modified \fewbody\, computing 10 realizations of each encounter while re-drawing the randomly selected parameters (e.g.~binary phase) to obtain a better statistical representation of the binary-single encounter population.

The resulting objects that leave the model to become runaway stars or HVSs are identified as follows.
We calculate the final velocity of an object after an encounter $v_{\rm fin}$ as the hyperbolic excess velocity
\begin{equation}
    v_{\rm fin} = v \sqrt{\frac{U + K}{K}},
\end{equation}
\noindent where $v$ is the velocity of the object leaving the encounter, and $U$ and $K$ are the Keplerian potential and kinetic energies of the top-level binary-single system (all of these quantities are evaluated at the termination of integration, which as noted above is a time where the potential may still have some significant magnitude).
The local escape velocity of the star cluster $v_{\rm esc}$ is provided by \CMC\ with the other encounter parameters; any object with $v_{\rm fin} \ge v_{\rm esc}$ is considered to escape the cluster.
The velocity of an ejected object once it has left the cluster is therefore
\begin{equation}
    v_{\rm out} = \sqrt{v_{\rm fin}^2 - v_{\rm esc}^2}.
\end{equation}

The initial conditions used for these encounters are calculated in the center-of-mass rest frame of the encounter, i.e. they do not contain information about the center-of-mass velocity of the encounter in the frame of the GC model.
We do not attempt to correct for this, which leads to a minor underestimation of final velocities for encountering objects and subsequently ejection rates (we justify this in Appendix \ref{app:restframe}).

\section{GC dynamics and BSCO ejections} \label{sec:gcdyns}

\subsection{Core collapse in GCs} \label{subsec:cc}

Core collapses are well-documented features of GC evolution \citep[e.g.][]{1968MNRAS.138..495L, 2001A&A...375..711F, 2006MNRAS.368..121F, 2008gady.book.....B, 2020IAUS..351..357K}, and the heightened densities during such phenomena makes them relevant to the present study.
This process occurs due to the energy transport by two-body relaxation from the central regions of the cluster outwards, causing a core to develop and contract while the outer regions expand \citep{2003gmbp.book.....H}.
The presence of massive objects (e.g. BHs) can significantly accelerate this process, as dynamical friction slows the most massive objects in the system, causing them to segregate into the core on a much more rapid timescale \citep{2008gady.book.....B}.
Following \cite{2013MNRAS.432.2779B}, we refer to this initial BH-lead collapse as the ``first’’ core collapse.
While this process can be effected substantially by the metallicity of the star cluster, with more metal-rich clusters experiencing greater mass loss due to stellar winds and producing lower-mass BHs with correspondingly longer dynamical friction timescales (fewer of which are retained due to natal kicks) \citep{2022arXiv220316547R}, core collapse typically occurs within a few 100 Myr for most GCs. 

% From the beginning of a GC's lifetime, dynamical friction slows the most massive objects in the system, while simultaneously accelerating the less massive objects \citep[e.g.][]{2008gady.book.....B}.
% Globally, this leads to the expansion of the outer layers of the cluster with a contemporary contraction of the core \citep{2013MNRAS.429.2881C}, usually within 100 Myr of the birth of the system.

%This first core collapse is delayed and weakened in systems with higher metallicities: the increased mass loss due to stronger stellar winds directly leads to smaller objects that experience weaker dynamical friction, and indirectly the number of ``massive" objects is reduced as any that to evolve into BHs are more easily removed from the GC by natal kicks \citep{2008MNRAS.386...65M, 2022arXiv220316547R}.

First core collapse is stopped by the formation and hardening of binaries in the core (which at this point is largely composed of BHs), and the liberated energy supports the universal expansion of the GC for a time on the order of Gyr \citep{2012MNRAS.420..309B}.
The BH population in the core is gradually depleted as strong encounters accelerate their participants beyond the local escape speed, until no more than a handful of BHs remain in the cluster \citep{2013MNRAS.432.2779B, 2020IAUS..351..357K}.
The loss of the dynamical heat source causes the core to contract again, as the system adjusts to being supported by WDs, NSs and stars in the same mass regime.
Unlike the first collapse, this second, permanent core collapse is observable in that the surface brightness profile of the GC becomes cuspy to the limit of the cluster center (and is what observers would classically call a ``core-collapsed cluster'').
We follow \citet{2018ApJ...855L..15K} in pairing these demographic and observable features, marking the time of second core collapse in our models as the time when the number of BHs drops below 10.

\subsection{Ejections from individual clusters} \label{subsec:single_clusters}

The 148 \CMC\ models are designated by cluster initial size $N$ (number of stellar objects), initial virial radius $r_{\rm vir}$ (parsecs), distance from the galactic center $r_{\rm gc}$ (kiloparsecs, used in \CMC\ to calculate tidal effects from the galactic potential), and metallicity $Z$ (used to prescribe star evolution), with values chosen to span much of the MWGC parameter space; see \citet{2020IAUS..351..357K} for more details.
To get a clear idea of how cluster evolution is linked to BSCO ejections, we examine four sample models; our base \CMCcat\ model has initial conditions
\begin{itemize}
    \item $N = 8 \times 10^5$
    \item $r_{\rm vir} = 0.5~{\rm pc}$
    \item $r_{\rm gc} = 8~{\rm kpc}$
    \item $Z = 0.01Z_\odot$.
\end{itemize}
Each of the other three sample models vary one of these parameters (either $N \to 4 \times 10^5$, $r_{\rm vir} \to 2~{\rm pc}$, or $Z \to Z_\odot$) to promote understanding of the effects of each.
The base $N$, $r_{\rm gc}$, and $Z$ were chosen as they correspond to typical values for MWGCs; the notably small $r_{\rm vir}$ is chosen for the samples because the dynamic range of cluster evolution and stellar ejections is greatest among \CMC\ models of this size, making it easier to identify the same features present in larger models.

\begin{figure*}
    \script{cmc_single_clusters.py}
    \gridline{
        \fig{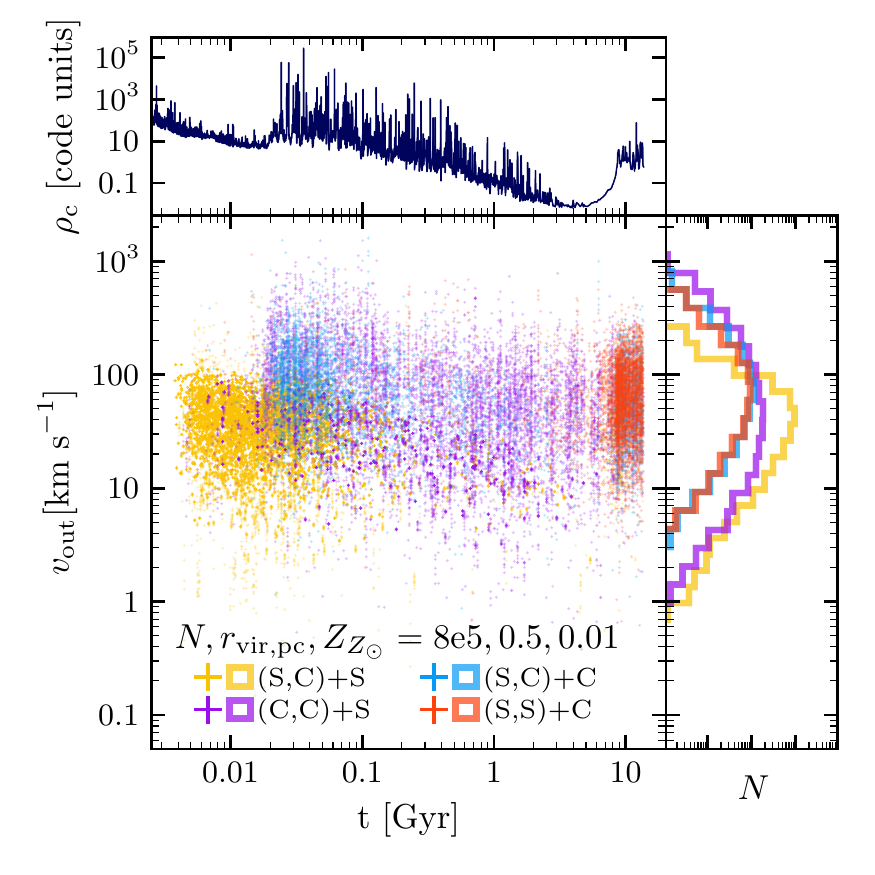}{0.49\textwidth}{}
        \fig{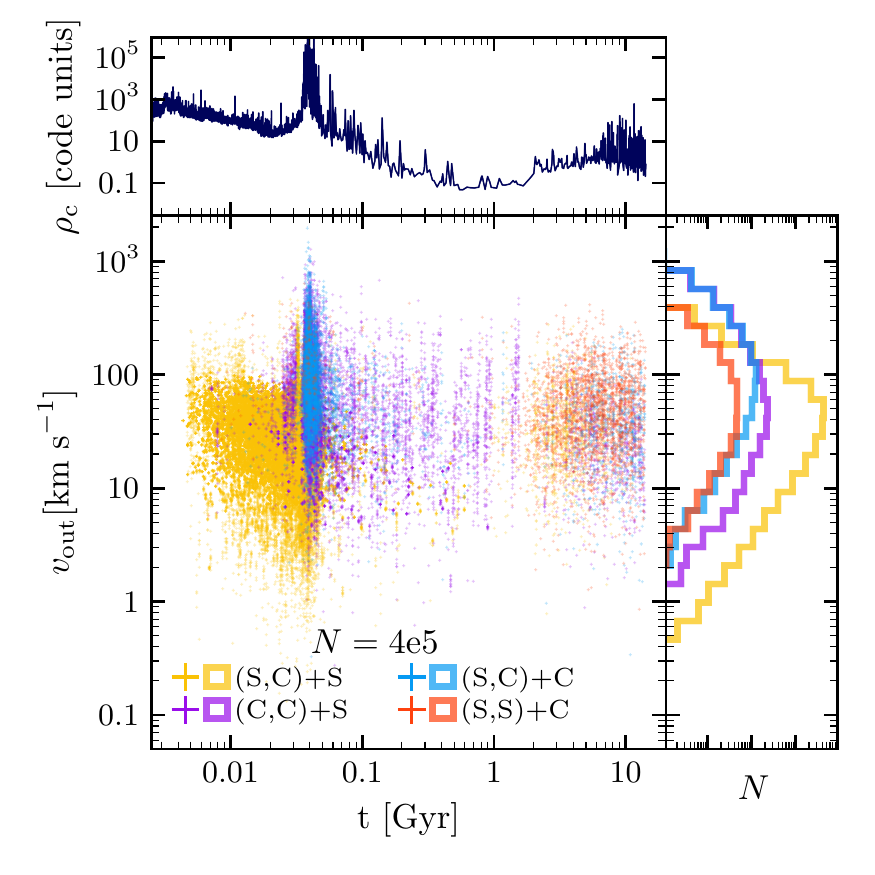}{0.49\textwidth}{}
    }
    \gridline{
        \fig{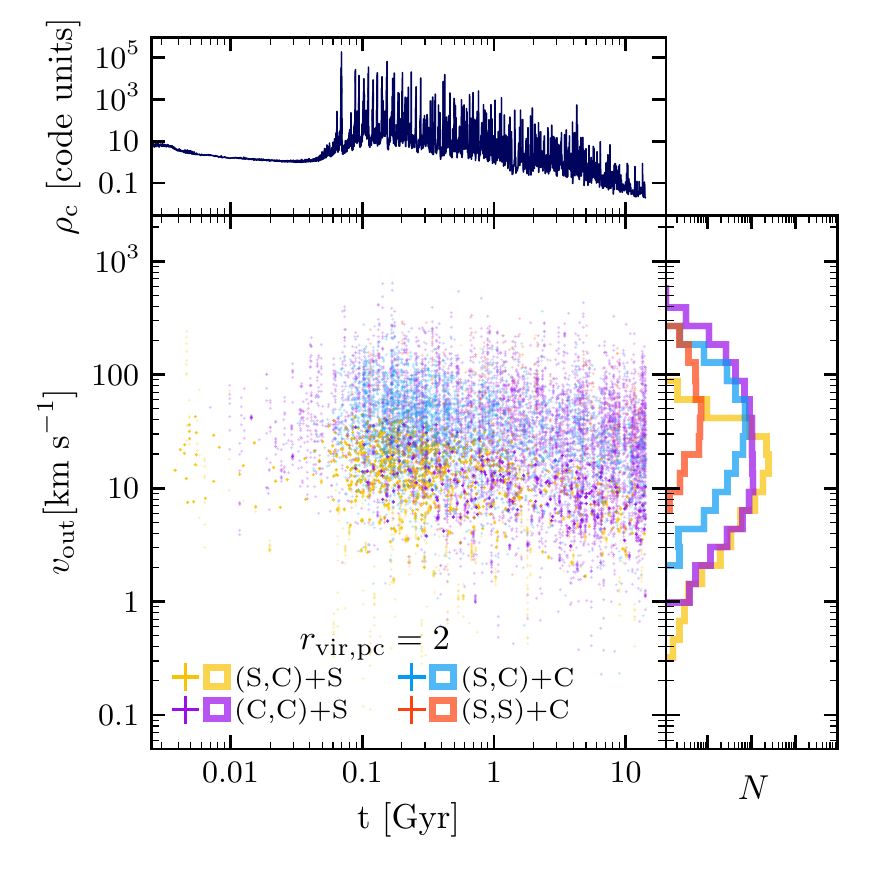}{0.49\textwidth}{}
        \fig{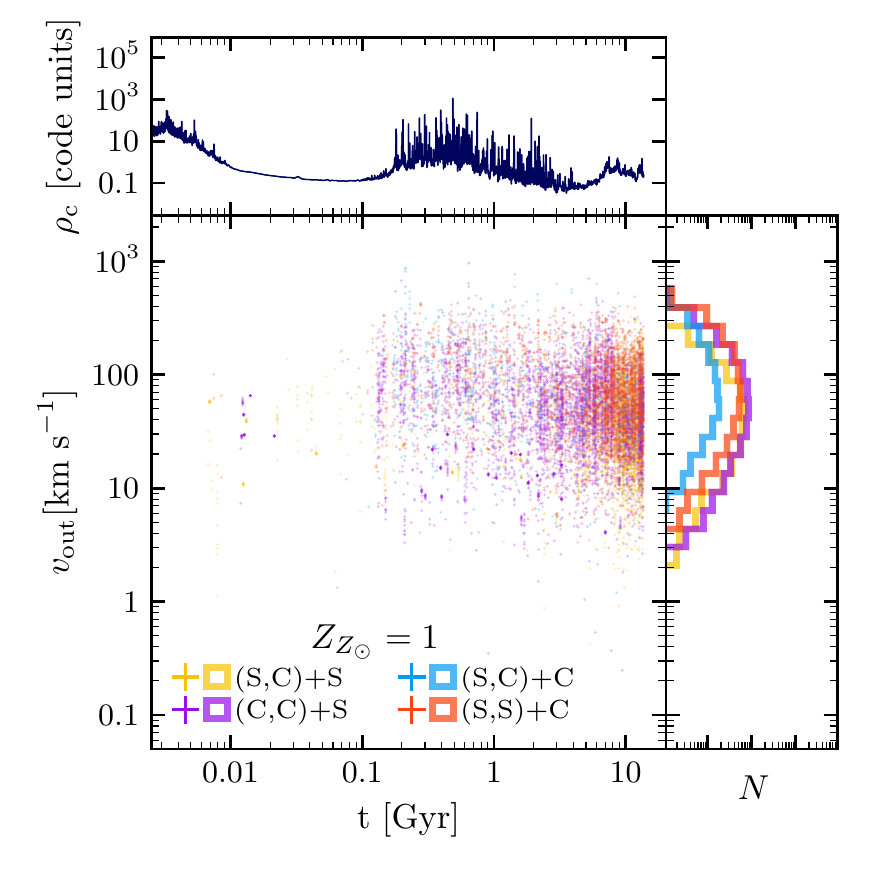}{0.49\textwidth}{}
    }
    \caption{
        Scatter plots of the cluster ejection velocity $v_{\rm out}$ versus encounter time $t$ for every escaping object from the integrated encounters for the four sample \CMC\ models (see the beginning of \S\ref{subsec:single_clusters} for details); the histograms show the distribution of velocities.
        The points are color-coded by the kind of encounter they originated from: encounters between a binary star and a CO are in red, encounters between a mixed binary (1 star and 1 CO) and a CO are in blue, encounters between a mixed binary and a star are in yellow, and encounters between a CO binary and a star are in purple.
        The core density (in code units) is plotted in above the scatter plot.
    }
    \label{fig:cmc_single_clusters}
\end{figure*}

Figure \ref{fig:cmc_single_clusters} shows the ejection velocity $v_{\rm out}$ for every MS star ejected through a BSCO encounter in these models, plotted at the time of encounter from initialization $t$; generally, these ejecta account for $\sim$1-6\% of all stellar escapers from a model over its lifetime.
What is immediately apparent from these single cluster models is a strong correlation between BSCO encounters and the evolution of the cluster core: whenever a cluster undergoes a core collapse, the heightened densities lead to predictably high encounter rates.

An interplay between stellar evolution and cluster dynamics is revealed in the kinds of BSCO encounters that occur at certain times.
The most massive stars in a cluster are both the first to form binaries and the first to evolve into COs.
Hence, encounters between a mixed binary (composed of a MS star and a CO) and a single MS star have the first opportunity to contribute significantly to the overall BSCO encounter population; in the densest models, this primacy leads to a majority of BSCO encounters being of this type over the 14 Gyr evolution.
Multiple-CO encounters first become major players at first core collapse after CO formation, and can take over as the majority in the case of models with weak or nonexistent pre-core collapse BSCO encounter phases.
There is a stark difference in the ejection velocities produced by these multiple-CO encounters in comparison to the previous mixed binary-single MS star encounters, and the fastest ejections over the cluster's lifetime are produced in the first tens of Myr after the associated first core collapse.
This regime lasts for a few to tens of Gyr, during which the COs in the core (predominantly BHs) preferentially form CO-CO binaries, as evidenced in the drop of mixed binary-single CO encounters during this phase.

For cluster models that evolve to BH-depletion second core collapse (in our sample, all but the $r_{\rm vir} = 2~{\rm pc}$ model), the corresponding increase in BSCO encounters is dominated by stellar binary-single CO encounters.
We also find that the majority of these encounters do not involve BHs, but rather WDs (or in some cases a NSs).
These behaviors are consistent with mass segregation, where it is only after the higher-mass BHs are ejected that these lower-mass COs can migrate into the core.

This general evolution varies with model parameters, as can be observed by comparing different panels of Fig. \ref{fig:cmc_single_clusters}.
The $N = 4 \times 10^5$ model lacks a core energetic enough to sustain a larger spatial profile, and the resulting contraction of the core leads to an ejection of 5\% of the cluster mass via BSCO encounters in the first 100 Myr.
This model does reach a BH-ejection core collapse by the end of the integration time, but this collapse is less pronounced than that for the base model.

The $r_{\rm vir} = 2~{\rm pc}$ model evolves to first core collapse later than the previous two, and noticeably lacks the pre-core collapse abundance of mixed binary-single star encounters seen above.
When first core collapse does occur, the distribution of encounters is similar to the respective distributions for the base model.

The difference between the solar metallicity model and the others is striking.
This model does not evolve to first core collapse until $\sim$200 Myr (by far the latest out of the four), and leads to typical densities no more than 10\% those of the other models; this occurs due to the lower BH masses and abundances associated with the greater mass loss (c.f. \S\ref{subsec:cc} above, and \S5.2 in \citet{2022arXiv220316547R}).
Also unlike the other models, here the number of ejections does not peak until the second, BH-depletion core collapse near 7 Gyr.

\begin{figure}
    \script{cmc_single_clusters_vesc.py}
    \centering
    \includegraphics[width=0.49\textwidth]{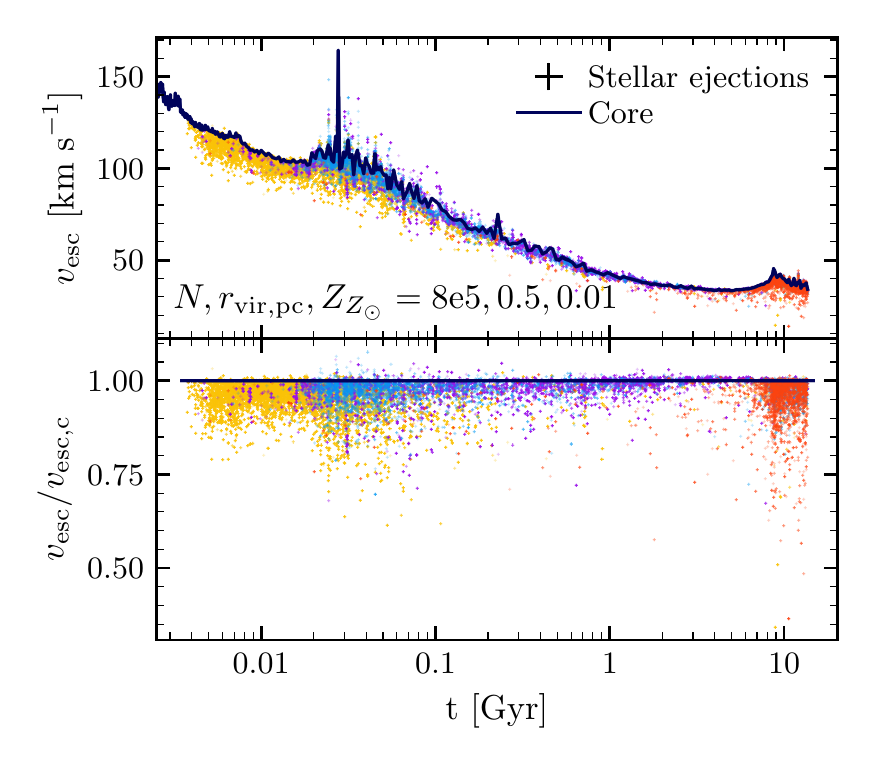}
    \caption{
        The local escape velocities $v_{\rm esc}$ at the locations of all BSCO encounters generated from the first sample \CMC\ model (colored points), plotted alongside the central escape velocity of the model (dark blue).
        The top plot shows the data in units of ${\rm km~s^{-1}}$, while the lower plot shows the same data normalized to the continuum of the core escape velocity (dark blue line in the upper plot).
        See Figure \ref{fig:cmc_single_clusters} for an explanation of the color scheme.
    }
    \label{fig:cmc_single_clusters_vesc}
\end{figure}

The dominance of the core in overall BSCO encounter production is even more clear when considering the radial localization of the encounters.
\CMC\ does not record the radius where each strong encounter occurred, but it does record the local escape velocity.
For comparison, the escape velocity from the center of the model can be computed from the central and tidal boundary potentials, which are recorded throughout the evolution of the cluster.
Figure \ref{fig:cmc_single_clusters_vesc} plots these data for the first sample cluster, where the distribution of encounter escape velocities is strikingly correlated to the central escape velocity, indicating the degree to which the core dominates these dynamics.
Throughout the evolution of the model, most BSCO ejections occur at or near the central escape velocity (the same plots for the other sample models also display this feature).
There is a small fraction of points that lie above the core velocity line, which we attribute to the difference in methods employed to calculate escape velocities in the ejection and core contexts.

\begin{figure*}
    \script{cmc_catalog.py}
    \centering
    \includegraphics[width=\textwidth]{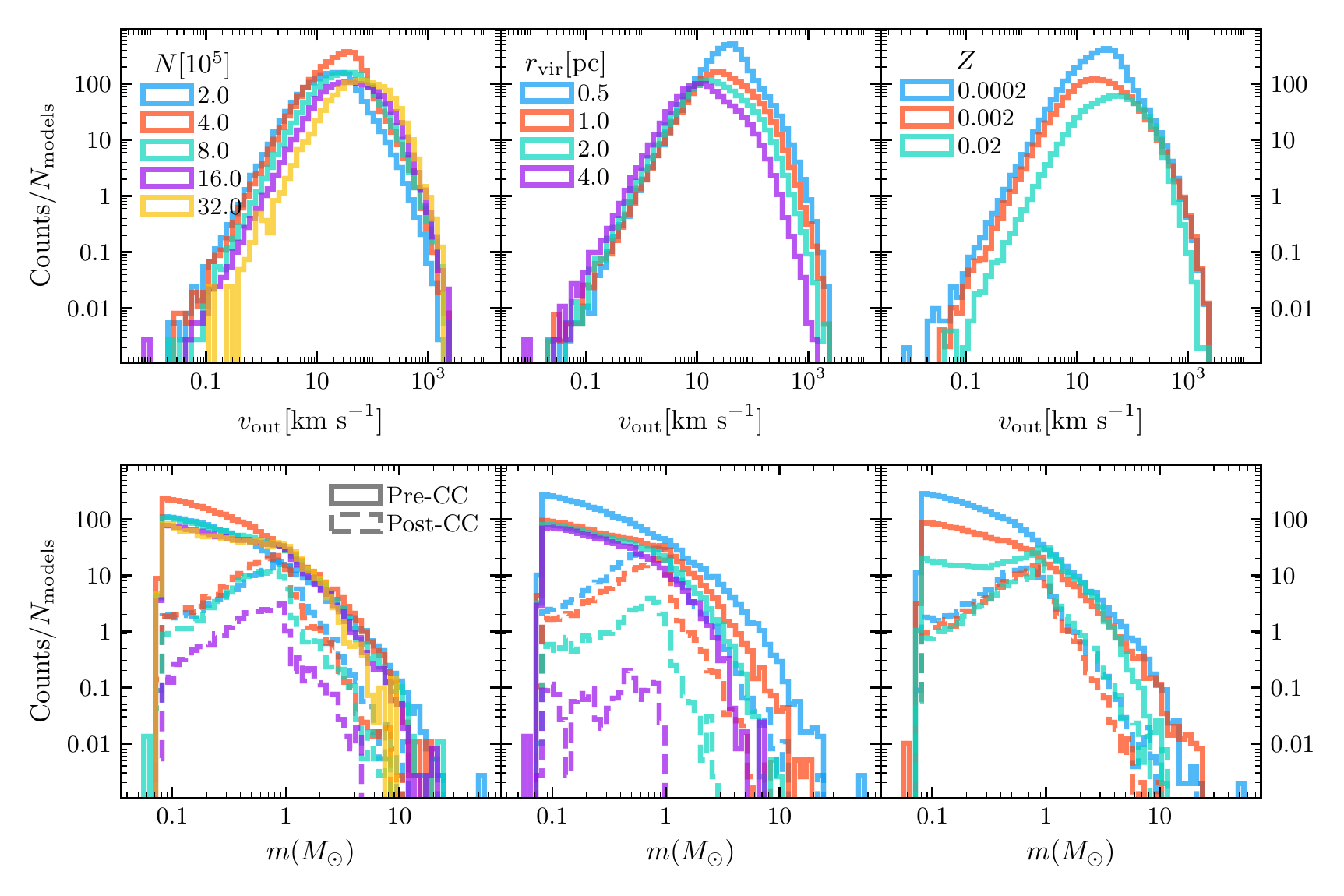}
    \caption{
        Histograms for all MS stars ejected from the \CMCcat\ models as a result of BSCO encounters.
        The top (bottom) row displays the distribution of ejection velocities from the models $v_{\rm out}$ (masses $m$ of the ejected objects).
        Each each separates the data by different \CMC\ model parameters: either size $N$ (number of objects), initial virial radius $r_{\rm vir}$ (parsecs), or metallicity $Z$.
        Each histogram is averaged over all models computed with the respective value of model parameter.
        In the the mass histograms, the data are further divided by whether the ejection occurred before or after the ``second," BH-depletion core collapse of the cluster, if one occurred within the integration time.
    }
    \label{fig:cmc_catalog}
\end{figure*}

Figure \ref{fig:cmc_catalog} shows histograms for cluster ejection velocities $v_{\rm out}$ and masses $m$ of all stars ejected by BSCO encounters in all \CMCcat\ models, binned by model parameters $N$, $r_{\rm vir}$, and $Z$.
The number of \CMC\ models corresponding to each parameter value varies, and so each histogram is averaged over the respective models, in addition to being divided by the factor of 10 in encounter multiplication.
It is worth noting the prominence of the $N = 4\times10^5$, $r_{\rm vir} = 0.5$ pc, $Z = 0.0002$ models in these averaged histograms: they are the cause of the peaks above the other distributions, as the three (one for each value of $r_{\rm gc} \in \{2, 8, 20\}$ kpc) produce significantly more ejections than the other models ($\sim$2.9, 2.0, and $1.4 \times 10^4$, respectively, versus the average number of $1-2 \times 10^3$ for the catalog).

Acknowledging these especially fecund models, predicable trends are visible in these histograms: increasing mass and decreasing size are both associated with higher ejection velocities.
However, while the number of ejections increases as the models become more compact, increasing the number of particles/mass of the model leads to a slight \textit{decrease} in ejecta.
The $N = 4 \times 10^5$ models consistently produce the most ejections in comparison with otherwise identical models of different population sizes, suggesting that at this mass there are not enough massive objects to prevent a severe initial core collapse through binary burning, but there are enough to facilitate a high number of BSCO encounters (the number of strong encounters in general is also maximized in the $N = 4 \times 10^5$ models).

The most noticeable distinction among models of different metallicities is the number of ejections, which decreases with increasing metallicity.
This is understandable, as in practice the lower metallicity models have greater BH populations than the higher metallicity models; quantitatively, the 1\% solar models have about $1.2 \times$ as many BHs at the time of first core collapse as the solar models, and maximum BH masses about $2 \times$ those of the same.
This weakens BSCO dynamics before and during first core collapse, as was first made visible in Figure \ref{fig:cmc_single_clusters}.
Because this effect takes place in the earlier stages of model evolution - namely, when mass segregation has had less time to separate lighter objects from the strong dynamics of heavier objects - this has the most significant effect of reducing the number of ejecta with $m < M_\odot$.
Notably, the different metallicity models have quite similar ejecta mass distributions after second core collapse, as the effect of metallicity on the masses of the remaining WDs and NSs is much less pronounced than on the now-ejected BHs.

\section{A MW-like population of BSCO ejecta} \label{sec:est_MW-like}

In the interest of predicting realistic statistics and rates for a MW-like GC population, we seek to assemble a synthetic MW-like population of GC-runaways from BSCO encounters.
The two steps involved here are 1) selecting representative \CMC\ models for galactic GCs, and 2) integrating the post-ejection orbits in the context of the MW to produce a present-day picture.

\subsection{Pairing \CMC\ models to MWGCs} \label{subsec:pairing}

We predominantly use the observational catalog of \citet{2018MNRAS.478.1520B} to obtain parameters for MWGCs.
This catalog lacks metallicity measurements for the objects; therefore we supplement with that of \citet{2010arXiv1012.3224H}.
12 of the GCs in the former catalog do not have metallicity measurements in the latter, leaving us with the 149 MWGCs we use in this analysis.

Assigning a representative model to each MWGC is nontrivial.
While many GCs are expected to be older than 12 Gyr, some could be as young as 9 Gyr \citep{2013ApJ...775..134V}; to reflect this, we find 11 timesteps as evenly spaced as possible between 10 and 13.5 Gyr for each \CMC\ model, yielding a set of model snapshots spanning the different \CMC\ initial conditions and sampling the models at different evolutionary stages.

\begin{figure}
    \script{cmcs-mwgcs_scatter.py}
    \begin{centering}
        \includegraphics[width=\linewidth]{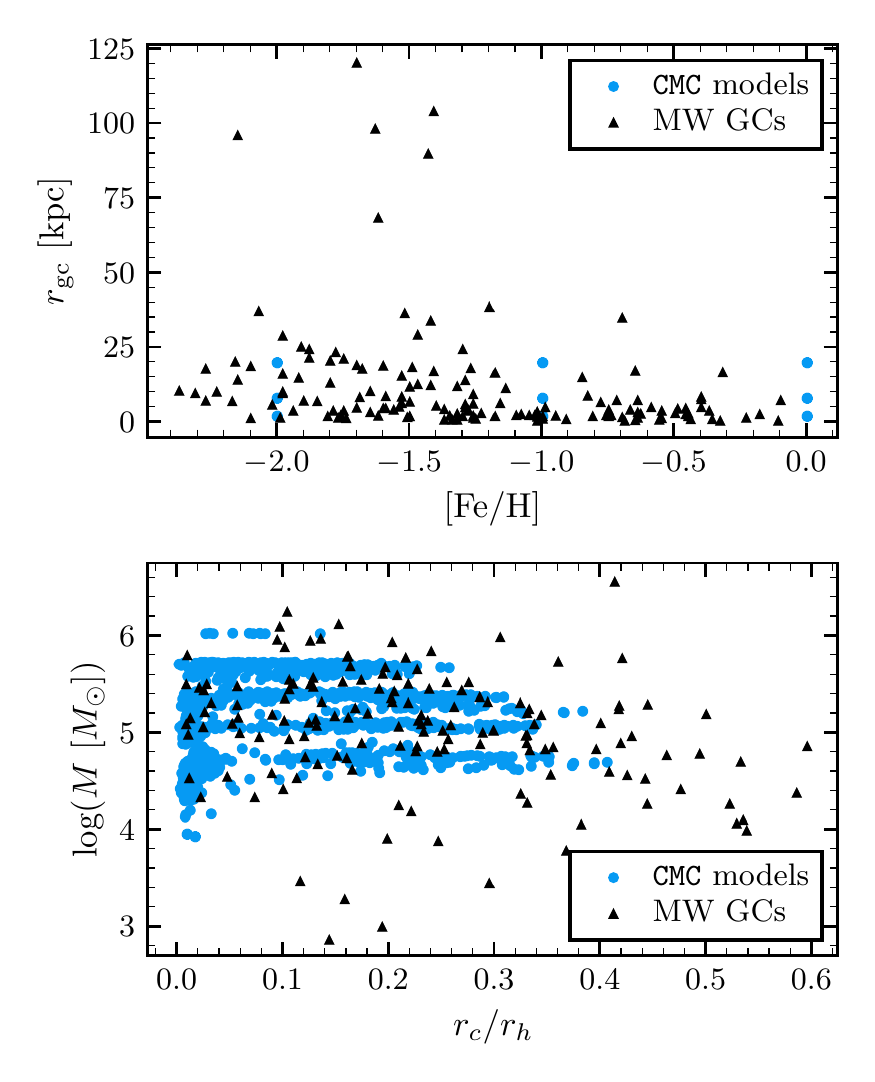}
        \caption{
        The clusters and models used in this work.
        Galactic GCs are represented by black triangles, and CMC model snapshots by blue dots.
        The top (bottom) plot shows the systems in $r_{\rm gc}$ v. [Fe/H] ($\log M$ v. $r_c / r_h$) space.
        }
        \label{fig:cmcs-mwgcs_scatter}
    \end{centering}
\end{figure}

These snapshots are plotted with our composite catalog of MWGCs in Figure \ref{fig:cmcs-mwgcs_scatter}.
The upper plot shows the models/GCs in $r_{\rm gc}$ - [Fe/H] space; note the discretization of the \CMC\ models, as the respective parameters are held constant over integration, and so each blue point represents the $\sim 15$ \CMC\ models that are initiated with those particular values.
When choosing a representative snapshot for each GC, we effectively follow \citet{2021ApJ...912..102R} by first finding the blue point closest to the GC in this space and constrain our search to the respective snapshots.

Having constrained the snapshots by $r_{\rm gc}$ and metallicity, we then choose one of an appropriate size and evolutionary state by comparing the snapshots with the model in normalized $\log M$ - $r_c / r_h$ space, where $M$ is the mass of the cluster in $M_\odot$, $r_c$ is the Spitzer core radius \citep{1987degc.book.....S}
\begin{equation}
    r_c = \sqrt{\frac{3 \sigma_c^2}{4 \pi \rho_c}},
\end{equation}
(where $\sigma_c$ is the central velocity dispersion and $\rho_c$ is the central density) and $r_h$ is the 3D half-mass radius of the snapshot (the latter two parameters are used because both are immediately accessible in the \citet{2018MNRAS.478.1520B} catalog and standard \CMC\ output).
The models and GCs are plotted in the non-normalized space of these parameters in the bottom plot of Figure \ref{fig:cmcs-mwgcs_scatter}.
The transformation to the normalized space is simply
\begin{equation}
    x_{\rm norm} = \frac{x - \bar{x}_{\rm \CMC}}{\sigma_{\rm x, \CMC}},
\end{equation}
where $\bar{x}_{\rm CMC}$ and $\sigma_{\rm x,CMC}$ are respectively the mean and standard deviation of the parameter $x$ over the complete set of \CMC\ snapshots.
The snapshot from the $r_{\rm gc}$ - [Fe/H] subset that is closest to the MWGC in the normalized $\log M$ - $r_c / r_h$ space is chosen to represent the GC; as an example, 47 Tuc ($r_{\rm gc} = 7.52$ kpc, $Z = 0.0019$; $\log M = 5.95$, $r_c / r_h,m = 0.125$ \citep{2018MNRAS.478.1520B, 2010arXiv1012.3224H}) is represented by a snapshot from model \texttt{N1.6e6\_rv2\_rg8\_Z0.002}, which at the 10.8 Gyr time of the snapshot has $\log M = 5.91$ and $r_c / r_h = 0.126$.

While the \CMCcat\ has fairly representative models for most of the MWGCs, there are a number of larger-cored GCs (right side of the bottom plot of Figure \ref{fig:cmcs-mwgcs_scatter}) that are relatively distant from the nearest \CMC\ model; in practice, this means that a single \CMC\ model can be used to represent several MWGCs.
In practice, we find that the models that are chosen for several MWGCs are average models for the \CMCcat, and so we expect that while characteristics of more extreme GCs are not necessarily well-represented, they are replaced by typical models nonetheless.

\subsection{Integrating runaways to the present day} \label{subsec:galpy}

Having chosen representative snapshots for all MWGCs, we now use our synthetic ejecta populations to compose a MW-like population of BSCO ejecta from GCs.
All orbit integrations in the galactic frame are done with \texttt{galpy}\footnote{http://github.com/jobovy/galpy} \citep{2015ApJS..216...29B}.
Setting the snapshot to the present-day galactic orbital parameters from \citet{2018MNRAS.478.1520B}, we back-integrate the orbit of each GC in the \texttt{galpy} \texttt{MWPotential2014} potential to the initialization time of the respective model.
We then locate each ejected object at the appropriate place in the GC orbit, using the time of BSCO encounter (naturally ignoring any ejections that occurred after the time of the chosen snapshot).
A random ejection direction is chosen in the GC rest frame for each object, and the velocities are then transformed to the galactic rest frame.
Finally, the orbits of all ejecta are then integrated to the present day in the same galactic potential.

\begin{figure*}
    \script{gc_orbit_ejections.py}
    \gridline{
        \fig{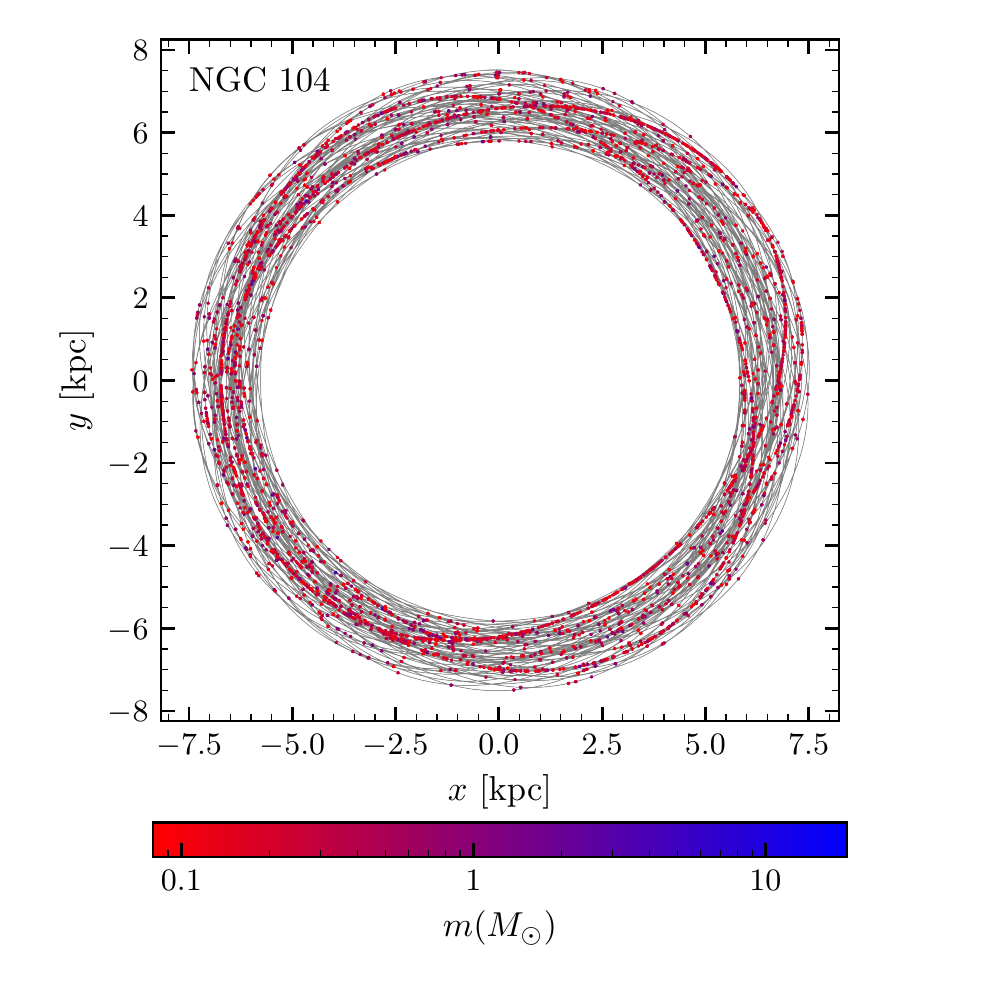}{0.49\textwidth}{}
        \fig{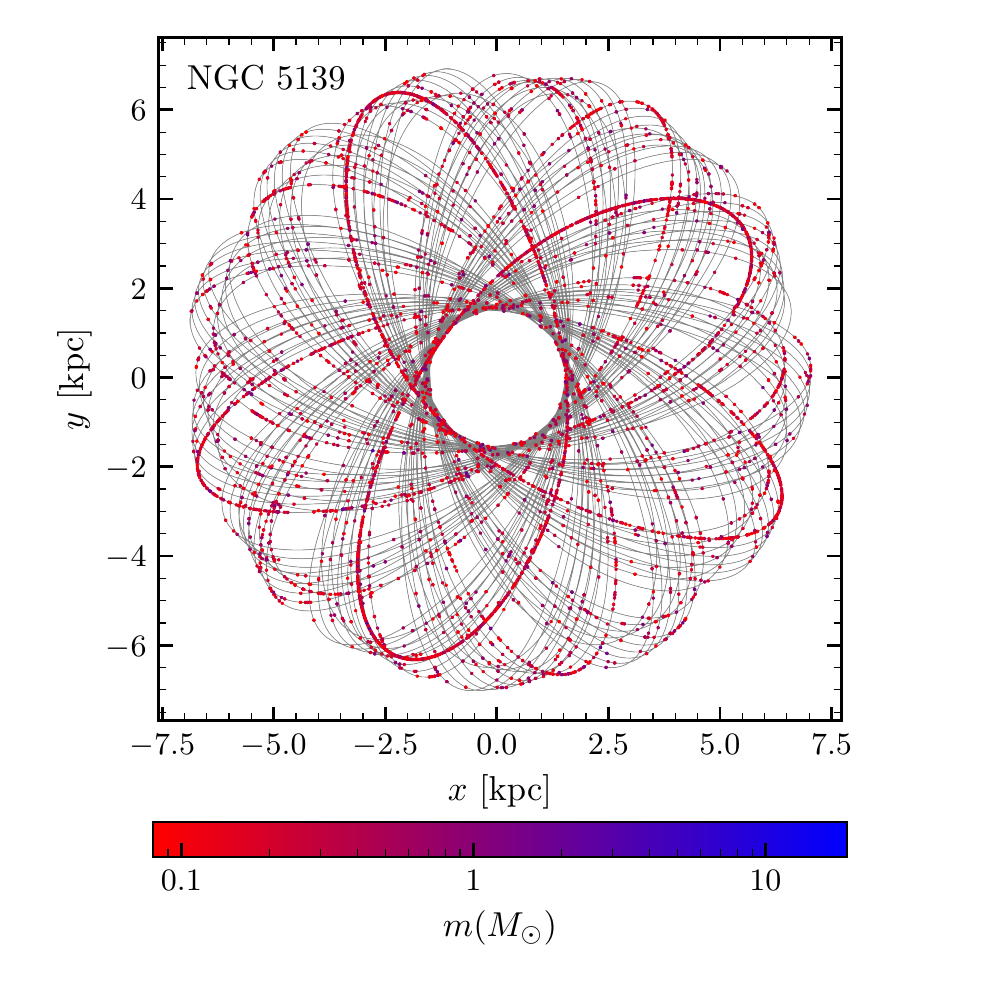}{0.49\textwidth}{}
    }
    \gridline{
        \fig{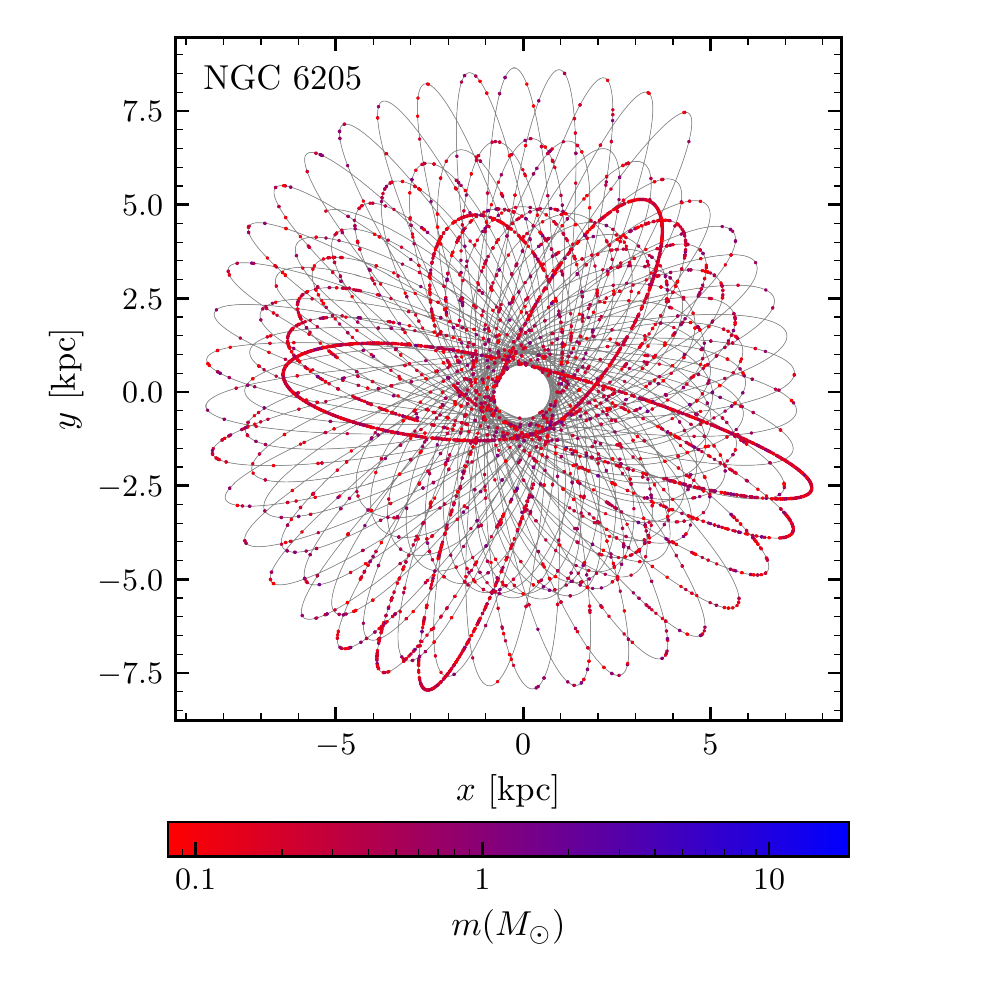}{0.49\textwidth}{}
        \fig{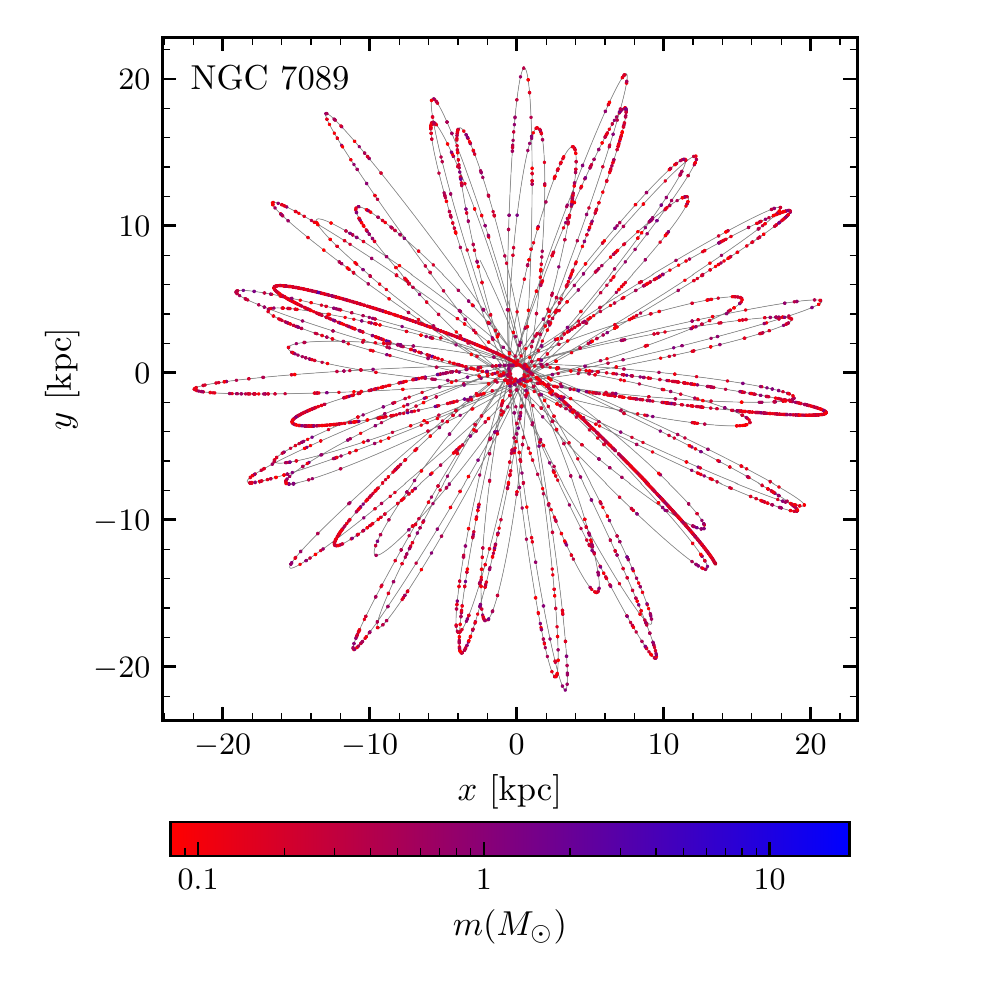}{0.49\textwidth}{}
    }
    \caption{
        Plots showing the back-integrated orbits for some sample MWGCs (gray curves), and the points in the orbit where an object is ejected from the GC (scatter points).
        The color scale communicates the mass of the ejected star.
        The concentration of ejections in the first few orbits is clear, and an increased density of ejections when clusters are farther away from the galactic center is visible as well.
        Viewing the figure electronically makes it easier to find the few high-mass ejecta amid the abundance of lower mass objects.
    }
    \label{fig:gc_orbit_ejections}
\end{figure*}

Figure \ref{fig:gc_orbit_ejections} shows the integrated orbits for a selection of GCs, and the points at which objects are ejected from the cluster.
As was seen earlier, ejections occur much more frequently in the early stages of the cluster, and here the slower velocities at larger distances from the galactic center leads to a higher percentage of objects being ejected far away from this focus.

\begin{figure*}
    \script{gcej_today.py}
    \gridline{
        \fig{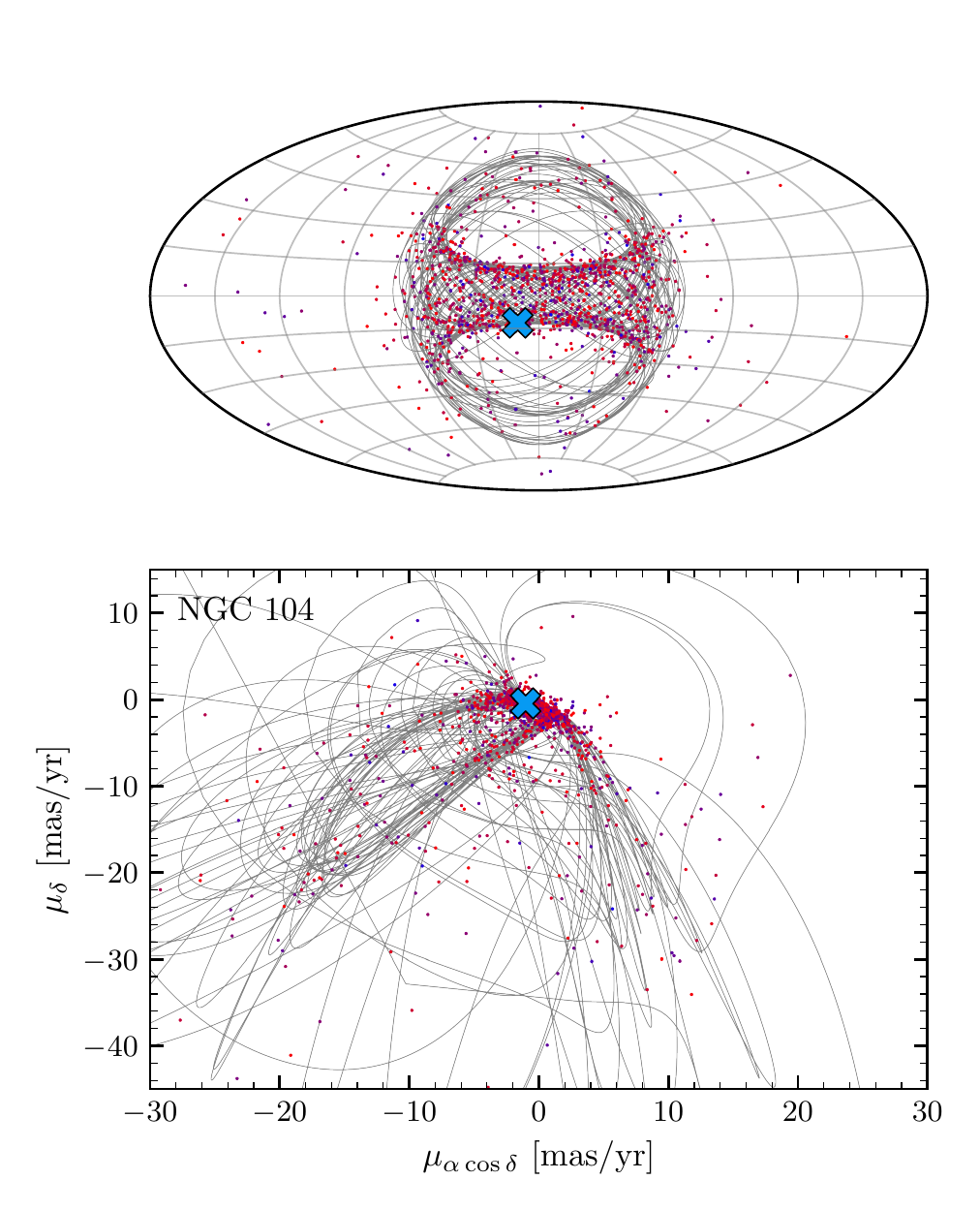}{0.425\textwidth}{}
        \fig{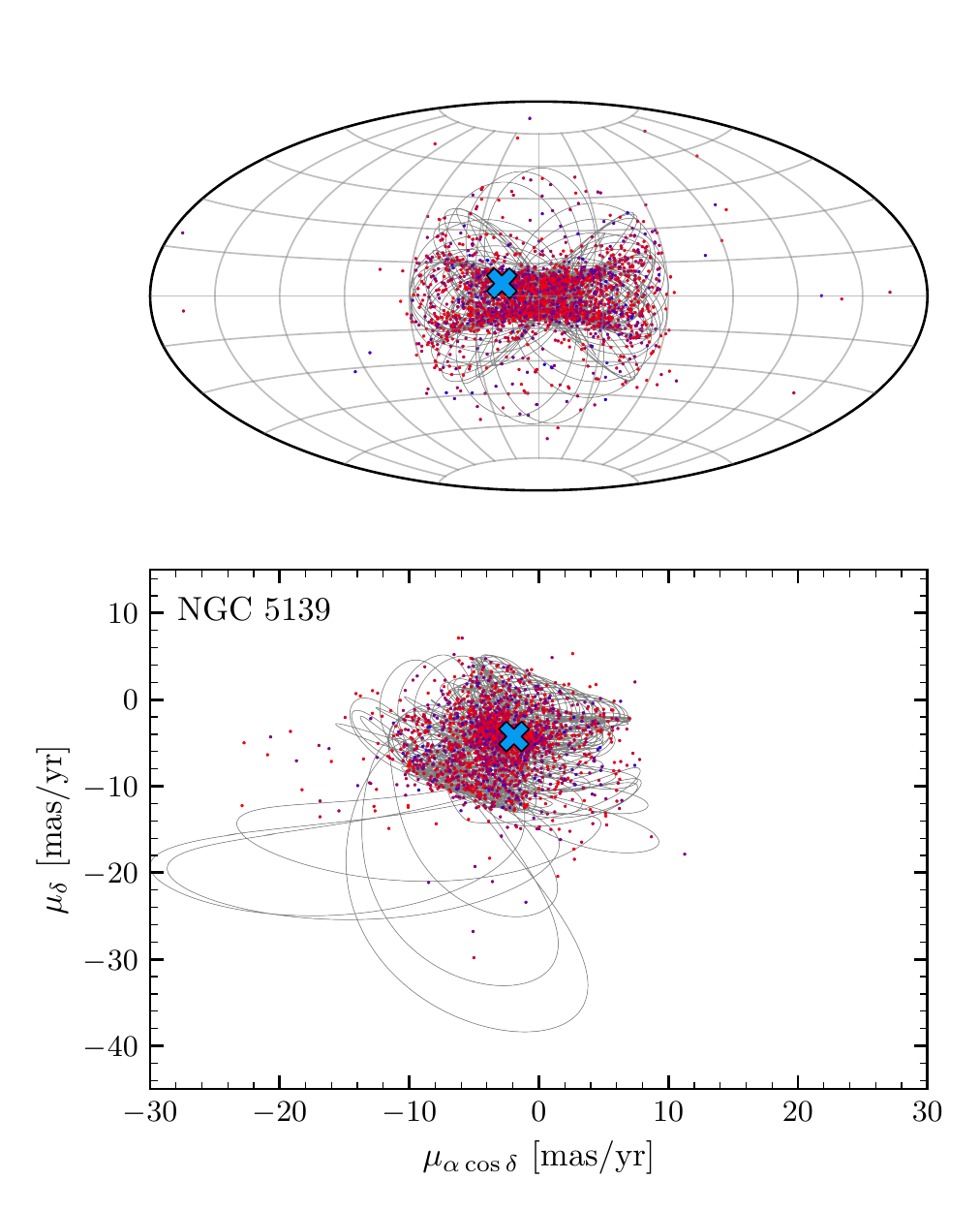}{0.425\textwidth}{}
    }
    \gridline{
        \fig{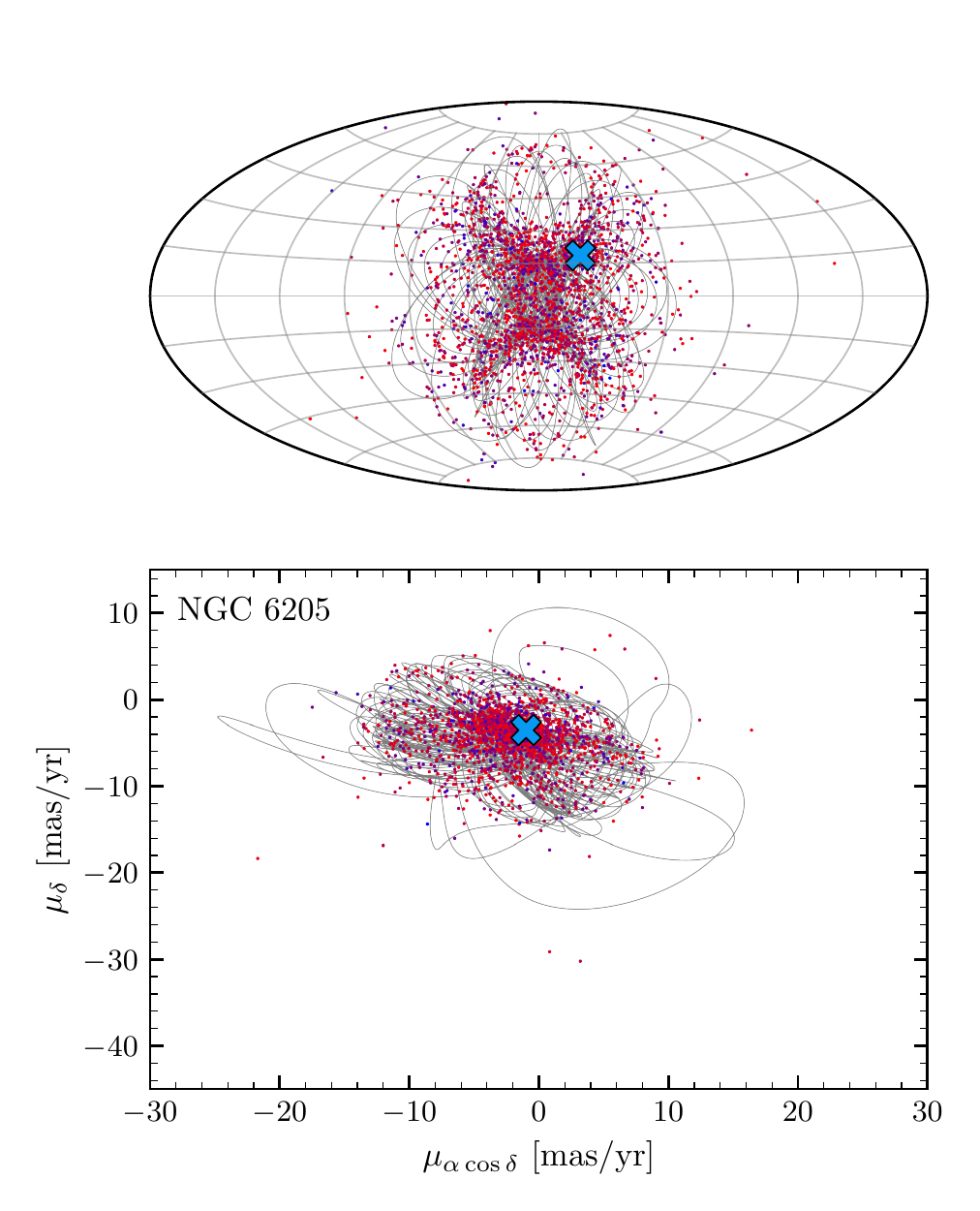}{0.425\textwidth}{}
        \fig{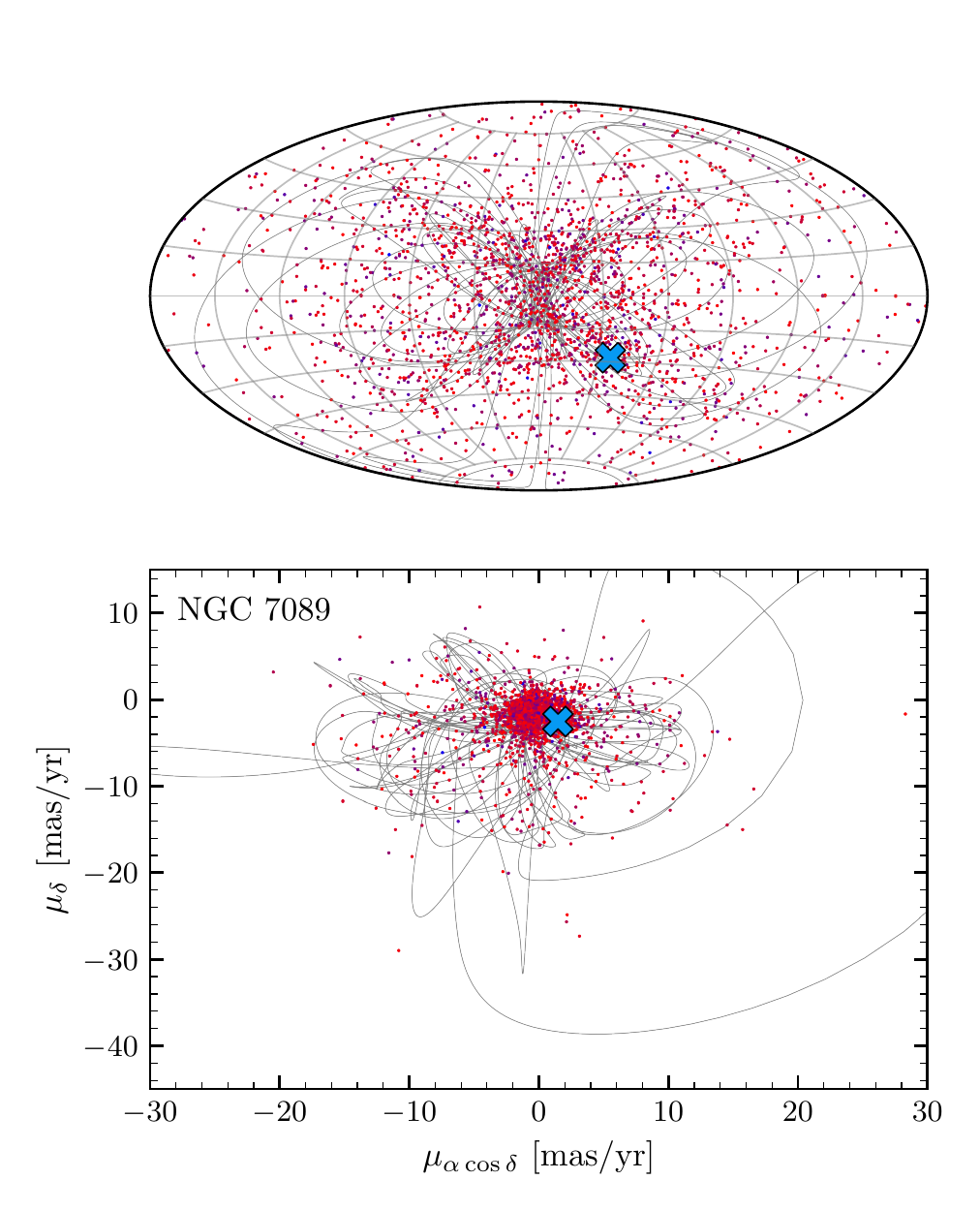}{0.425\textwidth}{}
    }
    \caption{
        Present-day positions (galactic longitude/latitude) and velocities (projected onto the galactic longitude/latitude directions) for the runaway objects from the sample GCs.
        The color scale is the same as in Figure \ref{fig:gc_orbit_ejections}, depicting the eject masses.
        The back-integrated orbits are shown as the gray trajectories, and the blue "x" is the position/velocity of the GC as measured by \citet{2018MNRAS.478.1520B}.
        The set of synthetic ejecta shown here is the result of downsampling the total set by a factor of ten, to account for the repeated-realizations method described in \S\ref{sec:methods}. 
    }
    \label{fig:gcej_today}
\end{figure*}

The present-day synthetic populations of runaways for the same sample CMC/GC pairings are shown in Figure \ref{fig:gcej_today}.
While it is clear how the GC orbits influence the distribution of ejected objects, the ejecta wander to a broad enough spread that any one object loses some of the information of its cluster of origin by the present day.
The proper motion distribution is less affected in this way (especially for more circular GC orbits), which is coherent with previous works that use these velocities to study the origins of such objects.
What is important to note here is that the current proper motion of the cluster is not necessarily the best locus to use when comparing stellar proper motions: a well-informed back-integration of the orbit can reveal the average proper motion of the cluster, which may be distinguishable from its present-day proper motion.
We include histograms, and 50\% and 90\% credible regions in sky area and proper motion space as a product of this work, for use as evidence when assigning runaway stars to GC origins; see Appendix \ref{app:credreg} for a description of these products.

\begin{figure*}
    \script{rgc-z.py}
    \begin{centering}
        \includegraphics[width=0.9\linewidth]{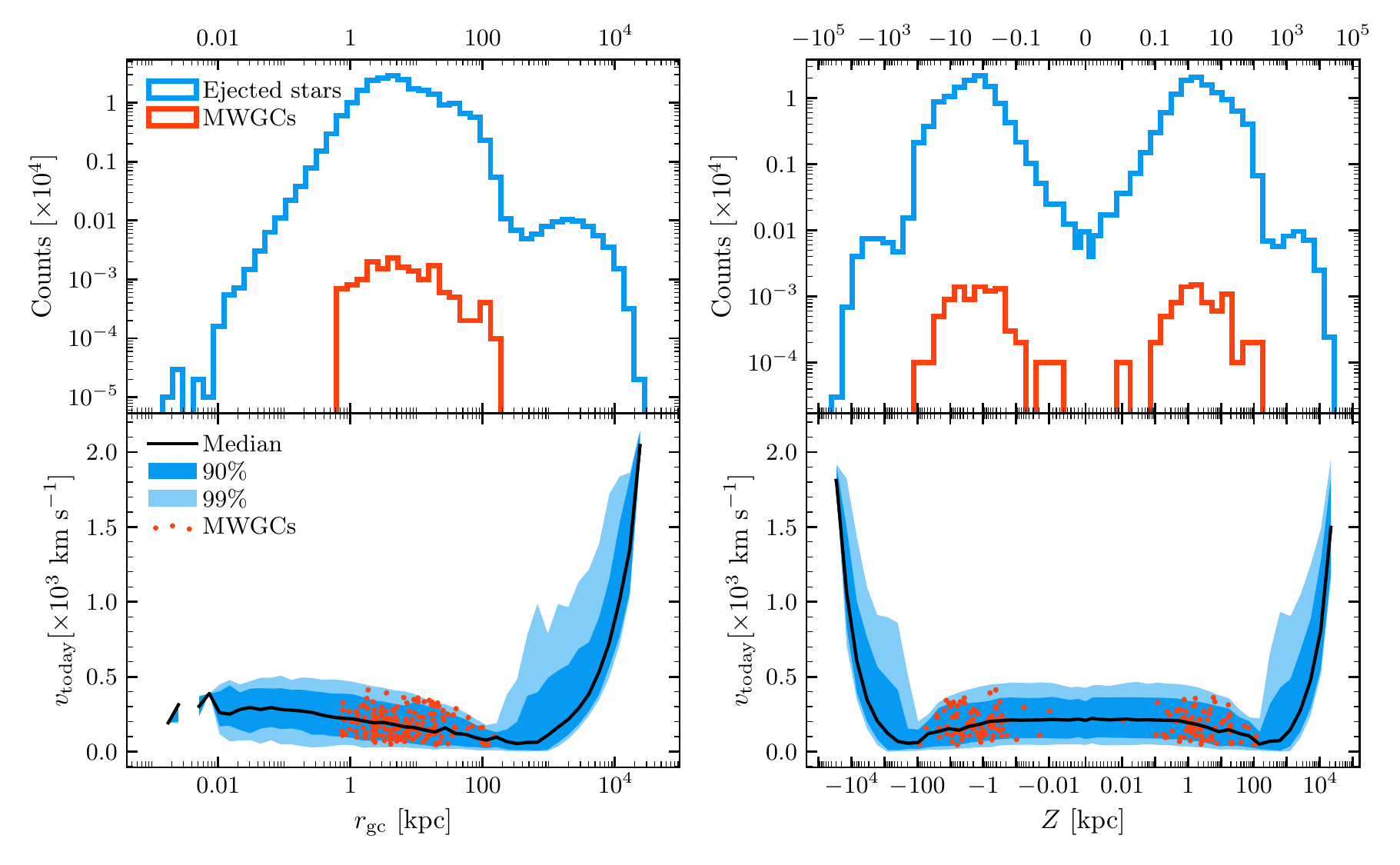}
        \caption{
            Histograms (top) and velocity quantiles (bottom) for the synthetic stellar ejecta.
            The left (right) plots show the profile over radial distance from the galactic center $r_{\rm gc}$ (distance from the galactic plane $Z$).
            The quantiles in the lower plots are calculated from the present-day velocities of our population.
            The red dots show the MWGCs from our composite \citet{2018MNRAS.478.1520B}+\citet{2010arXiv1012.3224H} catalog.
        }
        \label{fig:rgc-z}
    \end{centering}
\end{figure*}

From a galactocentric perspective, the population of runaways is fairly isotropic in position, as can be seen in Figure \ref{fig:rgc-z}.
The distributions of distance from the galactic origin, and the local distributions of galactocentric velocity are similar when distributing over radius $r_{\rm gc}$ versus distance from the galactic plane $Z_{\rm gc}$.
Most runaways end up at a distance on the order of 10 kpc from the galactic center, and with a velocity on the order of 100-300~\kms in the galactocentric rest frame.
Few runaways make it past the 100 kpc mark, but those that do naturally retain the highest velocities, reaching upwards of 1000~\kms.

\begin{figure}
    \script{cmc_orbits_cdf.py}
    \begin{centering}
        \includegraphics[width=0.99\linewidth]{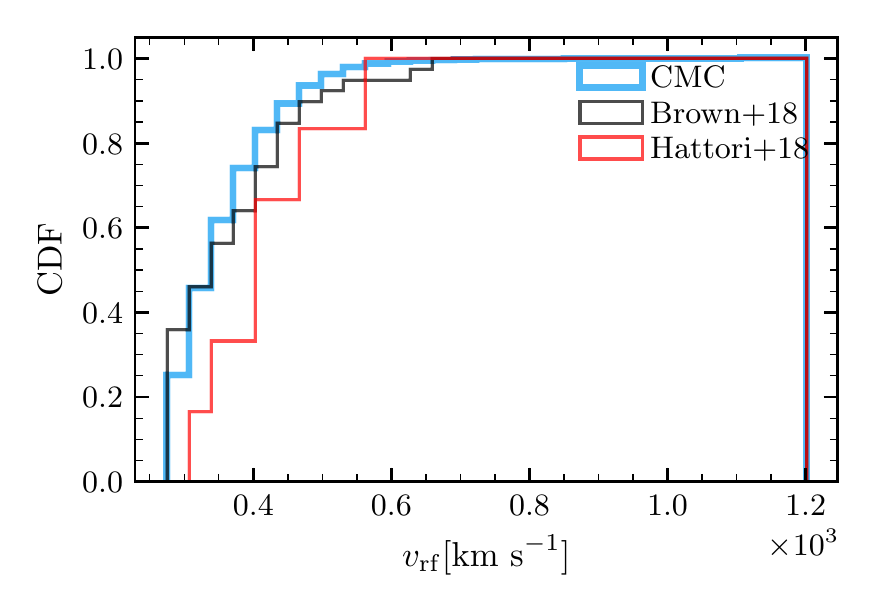}
        \caption{
            Velocity distributions for our synthetic and some previous observational HVS catalogs.
            $v_{\rm rf}$ is the heliocentric radial velocity of the star as measured in the galactic rest frame.
        }
        \label{fig:cmc_orbits_cdf}
    \end{centering}
\end{figure}

\begin{figure}
    \script{mwej_rates.py}
    \begin{centering}
        \includegraphics[width=0.99\linewidth]{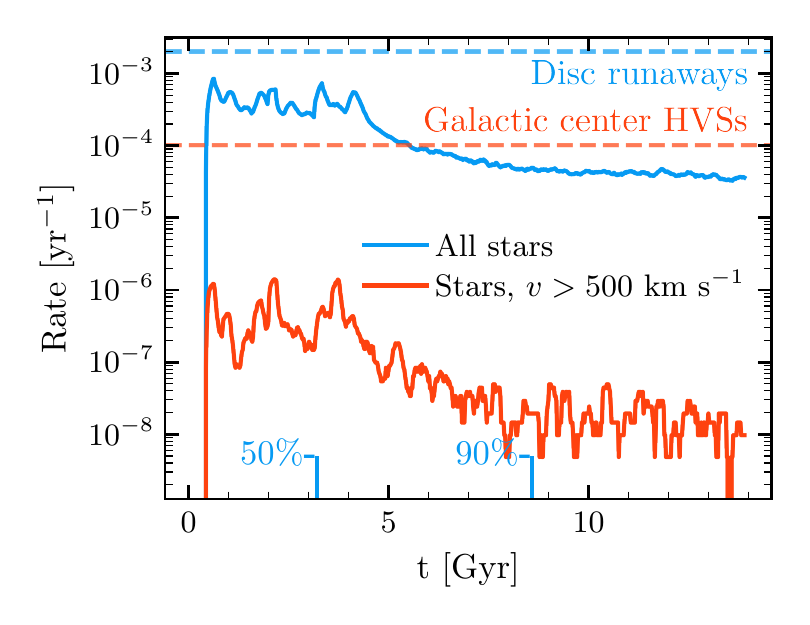}
        \caption{
            The evolution of the ejection rate for different components of our synthetic population.
            The blue (red) dashed line shows the estimated from observations for runaways from the galactic disc (HVSs from the galactic center), as reported by \citet{2015ARA&A..53...15B}.
            The threshold times by which 50\% and 90\% of the runaway ejections have occured are indicated along the bottom of the figure.
        }
        \label{fig:mwej_rates}
    \end{centering}
\end{figure}

One result of note is that of the heliocentric radial velocity, $v_{\rm rf}$, distribution of the synthetic population.
\citet{2021arXiv211213864G} in part studied the galactic center origin of HVSs, and found that there is an apparent tension between the observed and predicted runaway populations; specifically, the predicted number of stars with velocities $\gtrsim 700~{\rm km~s^{-1}}$ was much higher than the observed rate from the HVS sample of \citet{2018ApJ...866...39B}.
The same work found that a decreasing star formation rate over last millions of years would reduce the recent (and thus more likely to be observed) HVS production rate enough to overcome the discrepancy.
GCs naturally follow this pattern (c.f. \S\ref{sec:gcdyns}), as the BSCO runaways studied here are produced at much higher rates when the models are young, and few HVSs are produced in the later stages of the cluster.
The resulting $v_{\rm rf}$ distribution is much closer to the observed distribution than that for the galactic center origin case, albeit skewed towards slightly smaller velocities (Figure \ref{fig:cmc_orbits_cdf}).
This latter difference grows when comparing to the HVS sample of \citet{2018ApJ...866..121H}, who focused on metal-poor stars, which are more akin to the objects that populate GCs.

Finally, Figure \ref{fig:mwej_rates} shows the HVS/runaway rates for our synthetic population, along with observational rate estimates from \citet{2015ARA&A..53...15B}.
We note that 50(90)\% of ejections occur before our synthetic population is 3.21(8.59) Gyr old, and that naturally the earliest ejections are most dependent on any time variability in the MW potential.
At these earlier times, BSCO ejecta could have made as much as 20\% (1\%) of all runaways (HVSs), during the times when GCs underwent their first core collapses.\footnote{Here, the cutoff of this phase is fairly abrupt because we constrain our models to present-day ages of 10-13.5 Gyr (\S\ref{subsec:pairing}).  The physical rate will naturally be characterized by the evolution of the GC formation rate in the past.}
Afterwards, the ejection rates decay to the present day, where they are closer to 1.5\% (0.0001\%) of the observational values.
It is important to note nonetheless that GCs are able to produce HVSs with speeds high enough to make them comparable to those accelerated by mechanisms in the galactic center.

\section{Discussion \& Conclusions} \label{sec:disccon}

In this work we have studied strong 3-body encounters in GCs as means of producing stellar runaways.
We composed a synthetic MW-like population of ejecta by matching observed GCs to realistic $N$-body models of these systems and embedding the models in the orbits of their real counterparts.
In particular we considered binary-single encounters involving at least one compact object; this selection includes about half of all strong encounters in the catalog of models.

BSCO ejections were found to be closely linked to the evolution of the cluster core, where the closest encounters among the densest stellar objects occur.
GCs lose mass and expand as they evolve; accordingly, the majority of and the fastest BSCO ejecta were produced in the early stages of the models.
High-metallicity models had overall weaker ejection mechanisms (in frequency and maximum velocity), due to smaller stellar masses.
BSCO encounters occurring in realistic GCs are capable of accelerating stars to velocities in excess of 2000 \kms, which complicates the identification of ejection mechanisms for HVSs when their origins are not easily recognizable.
We also note that these velocities appear to pass the speed limit on star-only encounters found by \citet{1991AJ....101..562L}; further study is warranted to discern the impact of compact objects on small-$N$-body dynamics.

While ejected objects evolve to be largely indistinguishable from other MW stars in terms of position, they were found capable of retaining some information about the motion of their GC of origin, particularly in the case of GCs with near-circular orbits.
The overall population of ejecta was usually concentrated around the average proper motion of the GC throughout its orbit.
It is important to recognize that the present-day proper motion of a GC may not reflect this average proper motion, and that in general a better kinematic picture is accessible through back-integration of the orbit.

In the galactic context, the velocity distribution of the synthetic ejecta was found to be similar to that of HVS observations for velocities $\lesssim 500~{\rm km~s^{-1}}$; specifically, our population was skewed towards these velocities with respect to observations.
With galactic-center origin studies finding distributions skewed towards higher velocities in the same respect \citep{2018ApJ...866...39B}, it is possible that a mixture of the two could be used to more accurately model the real population of these objects.
Such a calculation must be done in light of the relative rates of the two mechanisms: our study concludes that while the GC BSCO runaway rate might have been a few 10\% of the overall rate in the first few Gyr of the MW, in the present day it is likely no more a few 1\% of the same.

\begin{acknowledgements}
This work was supported by NSF Grant AST-2009916 and NASA ATP Grant 80NSSC22K0722.
CR also acknowledges support from a Charles E.~Kaufman Foundation New Investigator Research Grant, an Alfred P.~Sloan Research Fellowship, and a David and Lucile Packard Foundation Fellowship.
\end{acknowledgements}

\software{\CMC\ \citep{2000ApJ...540..969J, 2013ApJS..204...15P, Rodriguez2022}, \fewbody \citep{2004MNRAS.352....1F}, \texttt{galpy} \citep{2015ApJS..216...29B}, \texttt{astropy} \citep{2013A&A...558A..33A, 2018AJ....156..123A, 2022ApJ...935..167A}, \texttt{numpy} \citep{harris2020array}, \texttt{pandas} \citep{mckinney-proc-scipy-2010, reback2020pandas}, \texttt{matplotlib} \citep{Hunter:2007}}

This work was prepared in part with the reproducibility software \href{https://github.com/showyourwork/showyourwork}{showyourwork} \citep{2021arXiv211006271L}, which leverages continuous integration to transparently connect the paper to the publicly-available dataset and relevant code.
The GitHub repository for this project can be found at \url{https://github.com/tomas-cabrera/hvss-bsco}; the icons next to paper content link to the particular scripts used to generate that content.
The dataset is stored at \url{https://doi.org/10.5281/zenodo.7599870}, which also includes copies of the simulation and figure scripts from the GitHub for insurance of future access.

\appendix

\section{Velocities in encounter rest frames versus the GC rest frame} \label{app:restframe}

We justify here our claim that in neglecting the transformation from the encounter rest frame to the GC model rest frame we obtain slower - and subsequently fewer - ejections.

H\'enon's Monte Carlo method assumes spherical symmetry, and subsequently the orbit of a particle is characterized by its energy, total angular momentum, and radial position; radial and tangential velocity magnitudes are drawn from the orbit by weighting by the amount of time the object spends at each point in its orbit (see \S2.5 of \citet{Rodriguez2022}).
The sign of the radial velocity is randomly chosen as positive or negative with equal weighting, and when setting up a strong interaction an angle $0 \le \phi \le 2\pi$ between the tangential velocities of the two objects is chosen from a uniform distribution.
These two features ensure isotropy of either velocity with respect to the other, and of the center-of-mass velocity of the encounter with respect to the GC model rest frame $\vec{v}_{\rm cm|GC}$.

If we assume that the direction of the post-encounter object velocity in the center-of-mass rest frame $\vec{v}_{\rm f|cm}$ is isotropic with respect to $\vec{v}_{\rm cm|GC}$, then the average speed after boosting back to the model rest frame is
\begin{equation}
    \langle v_{\rm f|GC} \rangle
    = \int_0^{2\pi} \frac{d\phi}{2\pi} \sqrt{(v_{\rm f|cm} + v_{\rm cm} \cos \phi)^2 + (v_{\rm cm} \sin \phi)^2}.
\end{equation}
Naturally, for $v_{\rm cm|GC} \ll v_{\rm f|cm}$ this average post-boost speed approaches $v_{\rm f|cm}$, and in the limit $v_{\rm cm|GC} \gg v_{\rm f|cm}$ it approaches $v_{\rm cm|GC}$, which recall must be less than the initial speed of the other object in the encounter: in either of these cases, the speed augmentation caused by the boost does not favor faster or slower post-encounter velocities.
The maximum average ``acceleration" resulting from the boost occurs at the limit $v_{\rm cm|GC} = v_{\rm f|cm}$, where $\langle v_{\rm f|GC} \rangle \sim 1.3 v_{\rm f|cm} = 1.3 v_{\rm cm|GC}$.

The assumption of isotropy in calculating $\langle v_{\rm f|GC} \rangle$ is appropriate for resonant encounters wherein the dynamics are chaotic.
For flyby encounters where the objects travel on roughly hyperbolic trajectories, there is a preference for post-encounter velocities in the same direction as the initial velocity, if the impact parameter distribution is sufficiently expansive and weighted by the square of the parameter and the final velocities are marginalized over the angle that orients the plane of the 2-body encounter between the star and the binary center-of-mass.
The isotropic case is therefore a conservative limit on this flyby case, and predicts a greater boost-propagated speedup than if the calculation was done in full detail.

In summary, neglecting to return to the model rest frame after the \fewbody\ step for strong encounters does not favor faster or slower post-encounter speeds when the final speed of an object is much greater than or much less than the center-of-mass speed of the encounter in the model rest frame.
In the case that the final speed of an object is comparable to the same center-of-mass speed, the average post-transformation boost of 30\% is unlikely to significantly increase the number of ejections, as the two speeds are generally less than the maximum initial speed among objects entering the encounter.

\section{Ejecta credible regions} \label{app:credreg}

\begin{figure}
    \script{check_credible_regions.py}
    \centering
    \includegraphics[width=0.9\textwidth]{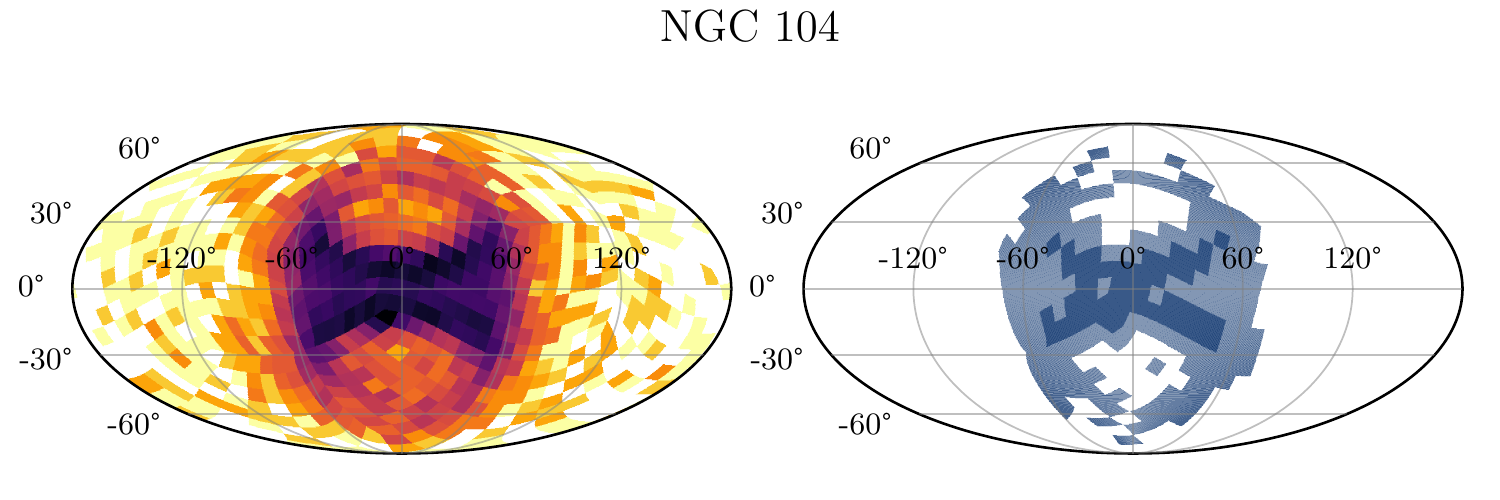}
    \caption{
        The HEALPix histogram for a sample GC (NGC 104), with the 50\% and 90\% credible regions in the right subplot.
        Galactic coordinates are used, and the histogram weights are log scaled.
    }
    \label{fig:check_credible_regions_x}
\end{figure}

\begin{figure}
    \script{check_credible_regions.py}
    \centering
    \includegraphics[width=0.7\textwidth]{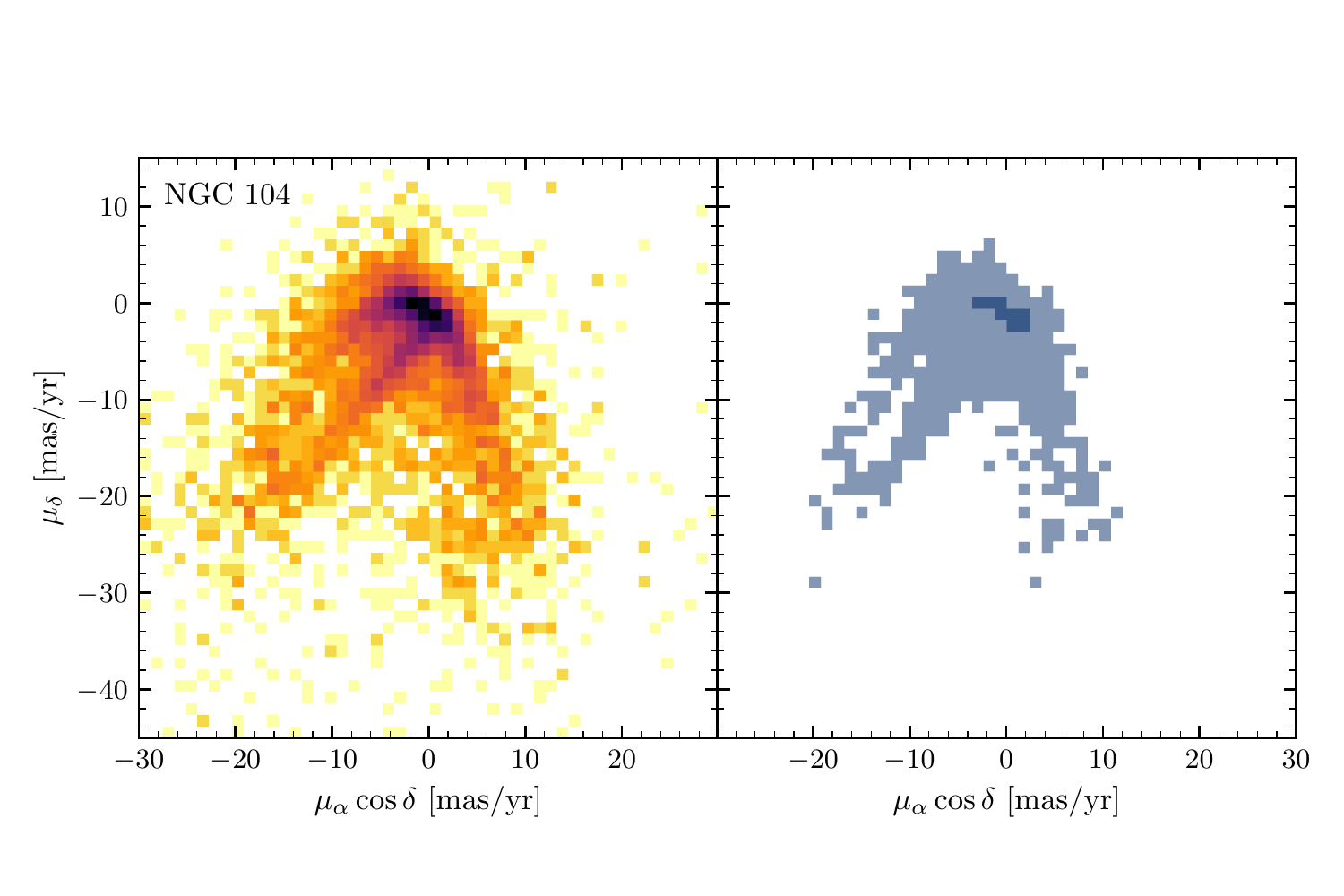}
    \caption{
        The proper motion histogram and credible regions for the same sample GC.
    }
    \label{fig:check_credible_regions_v}
\end{figure}

We include with this publication phase credible regions in phase space for ejecta populations from the MWGCs considered, which may be accessed at \url{https://doi.org/10.5281/zenodo.7599870}.
We intend these constructions to be used as secondary evidence for cluster membership of runaway stars, and to encourage further studies into understanding the respective phase space distributions.
The 2D position and proper motion distributions are separated into two different files, the conventions of which are described below.

The distribution of ejecta on the sky is quantified by discretizing the sphere with an order 4 nested HEALPix map \citep{2005ApJ...622..759G}, binning the sky into 3072 equal-area tiles.
We choose this resolution to enable identification of interesting credible regions while minimizing the apparent effect of isolated points whose exact location is dependent on the RNG seeds used.
We create the respective histogram by counting the number of synthetic ejecta that are found in each tile in the present day, and normalize by the total number of ejecta for the GC.
The resulting probability histogram is stored in the first column (\texttt{PROBS}) of the \texttt{hp\_probs.fits} file for each GC; the \texttt{NEJECT} field in the header of the same file contains the number of ejections for the GC.
We calculate our credible regions by cumulatively adding the highest probability bins until the target percentage is reached.
The last two columns (\texttt{CR50} and \texttt{CR90}) are boolean masks of the same convection as the HEALPix probability histogram corresponding to the credible regions (50\% and 90\%, respectively), where bins with entries of 1 are included in the region.
Figure \ref{fig:check_credible_regions_x} shows the histogram and 50\% and 90\% credible regions for NGC 104, as an example (the script used to generate this plot is \texttt{check\_credible\_regions.py}, which may be found in the GitHub/Zenodo repositories).

We construct proper motion histograms and credible regions in a similar manner.
We use a domain of $-30 \le \mu_\alpha \cos \delta~{\rm [km~s^{-1}]} \le 30$, $-45 \le \mu_\delta~{\rm [km~s^{-1}]} \le 15$ divided into a 50 $\times$ 50 grid.
These bounds and resolution are included in the headers of the \texttt{pm\_prob.fits} files for each GC, where calling \texttt{np.linspace(PM[D/RCD]MIN, PM[D/RCD]MAX, PMNUM)} will return the bin edges used for the appropriate dimension.
The header also includes a \texttt{COVERAGE} field containing the fraction of ejecta that lie in the specified domain; for all GCs this fraction is at least 0.98, and in most cases is greater than 0.999.
The total number of ejected objects (including those outside of the domain) is stored in the \texttt{NEJECT} field, as for the HEALPix histograms.
The proper motion histogram and respective masks for the 50\% and 90\% credible regions are stored in the \texttt{pm\_prob.fits} files as separate HDUs; these items for the same example GC as the HEALPix plot is show in Figure \ref{fig:check_credible_regions_v}, and the same \texttt{check\_credible\_regions.py} script contains the generating code.

\bibliography{bib}

\end{document}